\documentclass{amsart}
\usepackage[a4paper]{geometry}
\usepackage{amsmath,amssymb}
\usepackage[T1]{fontenc}
\usepackage[british]{babel}
\usepackage{graphicx}
\usepackage[utf8]{inputenc}
\usepackage[scr]{rsfso}
\usepackage{xspace}
\usepackage[sort]{cite}
\usepackage{tikz}
\usetikzlibrary{calc}
\usepackage{standalone}
\usepackage[inline]{enumitem}

\newcommand{\RR}{{\mathbb R}}
\newcommand{\scri}{{\mathscr I}}
\newcommand{\dd}{\mathbf{d}}
\renewcommand{\d}{\mathrm{d}}
\newcommand{\del}{\partial}

\newcommand{\etp} {\eth'}
\newcommand{\Y}[2]{{}_{#1}Y_{#2}}

\newcommand*{\sM}{\mathscr{M}}
\newcommand*{\tg}{{\tilde g}}
\newcommand*{\tildeh}{{\tilde h}}
\newcommand*{\tnabla}{{\widetilde\nabla}}
\newcommand*{\hnabla}{{\widehat\nabla}}
\newcommand*{\hgamma}{{\hat\gamma}}
\newcommand*{\tSigma}{{\widetilde\Sigma}}

\newcommand*{\CF}{\Theta}
\newcommand*{\cf}{\Omega}
\newcommand*{\tR}{{\tilde R}}
\newcommand*{\tC}{{\tilde C}}
\newcommand*{\tP}{{\tilde P}}

\newcommand*{\hP}{{\widehat{P}}}

\newcommand*{\ka}{{\textbf{a}}}
\newcommand*{\kb}{{\textbf{b}}}
\newcommand*{\kc}{{\textbf{c}}}
\newcommand*{\kd}{{\textbf{d}}}
\newcommand*{\ke}{{\textbf{e}}}

\newcommand*{\ub}{{\mathbf{u}}}
\newcommand*{\psib}{\pmb{\psi}}

\usepackage[colorlinks,linkcolor=blue,citecolor=red]{hyperref}
\DeclareMathOperator{\ArcTanh}{atanh}
\newcommand{\ee}{\mathrm{e}}

\newcommand{\Eqref}[1]{Eq.~\eqref{#1}}
\newcommand{\Eqsref}[1]{Eqs.~\eqref{#1}}
\newcommand{\Sectionref}[1]{Section~\ref{#1}}

\begin{document}

\title{Explorations of the infinite regions of space-time}

\author{Florian Beyer}
\address{Department of Mathematics and Statistics, University of Otago, PO Box 56, Dunedin 9054, New Zealand\\fbeyer@maths.otago.ac.nz}

\author{J\"org Frauendiener}
\address{Department of Mathematics and Statistics, University of Otago, PO Box 56, Dunedin 9054, New Zealand\\joergf@maths.otago.ac.nz}

\author{J\"org Hennig}
\address{Department of Mathematics and Statistics, University of Otago, PO Box 56, Dunedin 9054, New Zealand\\jhennig@maths.otago.ac.nz}

\begin{abstract}
An important concept in Physics is the notion of an isolated system. It is used in many different areas to describe the properties of a physical system which has been isolated from its environment. The interaction with the `outside' is then usually reduced to a scattering process, in which incoming radiation interacts with the system, which in turn emits outgoing radiation.

In Einstein's theory of gravitation isolated systems are modeled as asymptotically flat space-times. They provide the appropriate paradigm for the study of gravitational waves and their interaction with a material system or even only with themselves. In view of the emerging era of gravitational wave astronomy, in which gravitational wave signals from many different astrophysical sources are detected and interpreted, it is necessary to have a foundation for the theoretical and numerical treatments of these signals. 

Furthermore, from a purely mathematical point of view, it is important to have a full understanding of the implications of imposing the condition of asymptotic flatness onto solutions of the Einstein equations. While it is known that there exists a large class of asymptotically flat solutions of Einstein's equations, it is not known what the necessary and sufficient conditions at infinity are that have to be imposed on initial data so that they evolve into regular asymptotically flat space-times. The crux lies in the region near space-like infinity $i^0$ where incoming and outgoing radiation `meet'.

In this paper we review the current knowledge and some of the analytical and numerical work that has gone into the attempt to understand the structure of asymptotically flat space-times near space-like and null-infinity.
\end{abstract}

\maketitle

\section{Introduction}
\label{sec:introduction-1}

Einstein's field equation admits many exact solutions (see~\cite{Griffiths:2009,Stephani:2003}) most of which have only a very limited physical interpretation. However, there are two broad classes of solutions which are very important for our understanding of nature: these are the solutions describing cosmological models and the asymptotically flat solutions. We will be concerned here with the latter. Their importance stems from the fact that they describe the response of a gravitationally active isolated system under the impact of external fields such as electromagnetic or gravitational radiation or any other external matter fields.

The notion of an isolated system is an idealisation~\cite{Geroch:1977,Frauendiener:2004a,Frauendiener:2004} which tries to capture the essence of a system that is far from other massive bodies so that the interaction with its environment can be neglected. This implies that all fields, and in particular the gravitational field, must decay at sufficiently large distances from the system. That means that the curvature of the space-time describing the system vanishes and that the space-time becomes flat in the limit of infinite distances from the system.

In early works~\cite{Bondi:1962,Sachs:1962a,Sachs:1962b,Sachs:1962,Sachs:1961,Sachs:1960}  on the subject this characterisation of an isolated system was implemented by imposing suitable fall-off conditions on the metric or the curvature which were sufficient to guarantee the isolation of the system but at the same time also general enough to allow for interesting physical phenomena.

Proceeding in this way Bondi and coworkers~\cite{Bondi:1962,Sachs:1961} found a representation of the metric of an isolated system valid at large distances from the system which allowed them to prove the now famous \emph{Bondi-Sachs mass-loss formula} --- a theorem showing for the first time that Einstein's theory admits gravitational waves. They could demonstrate that the system loses energy, indicated as a reduction of the system's mass, by an amount which can be accounted for by a gravitational energy flux escaping to infinity.

At about the same time Penrose~\cite{Penrose:1964a,Penrose:1965,Penrose:1963,Penrose:1964} realised that the asymptotic fall-off conditions on the curvature that are imposed at infinity could be captured elegantly in a geometric way. He found that Einstein's vacuum equations, which are assumed to hold in the asymptotic regions of an isolated system, are almost invariant when the metric is rescaled by a conformal factor. Rescaling the metric corresponds to changing the units of space and time by the same factor which may vary from point to point. By choosing this factor appropriately one can achieve that the entire space-time has a finite size when measured in terms of the newly defined metric. Penrose realised that asymptotically flat space-times can be characterised by the fact that a conformal factor can be found so that the rescaled space-time admits a \emph{smooth} boundary. The points on this boundary represent events `at infinity', the fall-off conditions on the metric and the curvature turn into smoothness properties of the transformed geometry on the boundary. The main advantage of the conformal rescaling method is that questions about the global properties of space-times can be reformulated as questions about local properties of the conformal boundary as we will see repeatedly in this article.

The `re-interpretation' of asymptotically flat space-times as space-times with a smooth conformal boundary has proven to be an extremely successful step in the attempt to understand the properties of these space-times. One of the most important questions raised very early on concerned the size of this class of solutions. It was known by explicit construction that the Minkowski space-time, the Kerr-Newman class of black hole solutions and, more generally, all stationary space-times (which are naively perceived as asymptotically flat)  admit a smooth conformal boundary and are asymptotically flat in this sense. However, it was not clear whether any non-stationary space-time would fall into this class\footnote{As an aside we mention that the idea of conformal rescaling is not restricted to vacuum space-times nor to vanishing cosmological constant. For more details on these cases we refer to~\cite{Frauendiener:2004,Kroon:2016} and references therein.}. 

The next major advance in this area was the introduction of the \emph{conformal field equations} (CFE) by Friedrich~\cite{Friedrich:1981,Friedrich:1983a,Friedrich:1985,Friedrich:1986}, a formulation of Einstein's equation in the conformally rescaled context. The conformal field equations are a system of geometric partial differential equations (PDEs) whose solutions are completely equivalent to solutions of the Einstein equations. The only difference is that a `physical' space-time which satisfies Einstein's equations is obtained from the CFE as represented by an `unphysical' or `conformal' space-time together with a conformal factor which scales the unphysical metric to the physical metric. The CFE give access to the asymptotic properties of space-times without ever having to establish complicated limiting procedures. The equations have been used to prove several important theorems about the large scale structure and stability properties of asymptotically flat space-times, and in particular to provide a partial answer to the question about the size of the class of asymptotically flat space-times.

The conformal field equations have very similar properties as the Einstein equation: a $3+1$ decomposition of the space-time yields a system of constraint equations which is intrinsic to the space-like hypersurface used in the splitting and a system of evolution equations which can be made hyperbolic with an appropriate choice of coordinates. An important step towards understanding the asymptotically flat space-times was Friedrich's proof~\cite{Friedrich:1986} that the hyperboloidal initial value problem (IVP) for the conformal field equations is well-posed. A hyperboloidal hypersurface is a space-like hypersurface which, when described with the physical metric, stretches out to infinity like a hyperboloid i.e., for which the scalar curvature approaches a non-vanishing constant. In the unphysical space-time it is a space-like hypersurface which intersects the conformal boundary transversely. We will be concerned here only with hyperboloidal hypersurfaces intersecting \emph{future} null-infinity. Similar statements can be made about the past. Initial data satisfying the conformal constraint equations on such a hypersurface evolve toward the future into a piece of a space-time which satisfies the conformal field equations everywhere and is therefore a conformal rescaling of a solution of the Einstein equation. Friedrich went even further and proved that the evolved space-time is geodesically future complete provided that the initial data are close enough to Minkowski data. This means that the space-times evolving from such hyperboloidal data show the same global behaviour as the Minkowski space-time, they admit a smooth conformal boundary with a regular point $i^+$ at future time-like infinity. This theorem provides a semi-global stability result of Minkowski space-time under general small non-linear perturbations of initial data. It is semi-global because it applies only in the future direction.

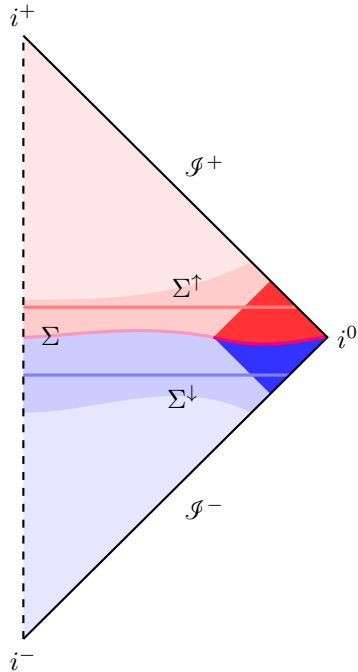
\begin{figure}
  \centering\begin{tikzpicture}[x=1cm,y=1cm]
  \coordinate (c1) at (0.5,0);
  \coordinate (c2) at (1.5,0.2);
  \coordinate (c3) at (3.25,-0.2);
  \coordinate (c4) at (3.75,0);
  \coordinate (p)  at (2.5,0);
  \coordinate (i0) at (4,0);
  \def\inner{ (0,0) ..controls (c1) and (c2).. (p)}
  \def\exact{ (p) ..controls (c3) and (c4) .. (i0)}
  \def\xxx{ (0,0) ..controls (c1) and (c2).. (p) ..controls (c3) and (c4) .. (i0)}
  \fill[red!10]  (0, 0.2) -- (3.8, 0.2) -- (0, 4) -- cycle;
  \fill[blue!10] (0,-0.2) -- (3.8,-0.2) -- (0,-4) -- cycle;
  \begin{scope}[fill=red!20]
    \fill [clip] \xxx -- (3.0, 1.0) ..controls(2, 0.5) and (1, 0.5) .. (0, 0.5) -- cycle ;
    \fill [red!80] let \p1 = (p) in ({\x1-2mm},{\y1-2mm}) -- (3.25,0.75) -- (i0) -- cycle;
  \end{scope}
  \begin{scope}[fill=blue!20]
    \fill [clip] \xxx -- (3.0,-1.0) ..controls(2,-0.5) and (1,-1.0) .. (0,-1.0) -- cycle;
    \fill [blue!80] (p) -- (3.25,-0.75) -- (i0) -- cycle;
  \end{scope}
  \colorlet{exactcolor}{magenta}
  \colorlet{innercolor}{magenta!50}
  \draw[color=innercolor,very thick] \inner node[black,pos=0.2]{$\Sigma$};
  \draw[color=exactcolor,very thick] \exact;
  \draw[very thick,red!50]  (0, 0.4) -- node[black,above,pos=0.6]{$\Sigma^\uparrow$} (3.6, 0.4);
  \draw[very thick,blue!50] (0,-0.5) -- node[black,below,pos=0.6]{$\Sigma^\downarrow$} (3.5,-0.5);
  \draw[thick] (0,4) node[above]{$i^+$} -- node[above right]{$\scri^+$} (4,0) node[right]{$i^0$} -- node[below right]{$\scri^-$} (0,-4) node[below]{$i^-$};
  \draw[thick,dashed] (0,-4)  -- (0,4) ;
\end{tikzpicture}
  \caption{A global asymptotically flat space-time constructed from Corvino-Schoen data}
  \label{fig:globalspacetime}  
\end{figure}

Several years ago, Corvino~\cite{Corvino:2000} and Schoen~\cite{Corvino:2006} proved a remarkable result: there exist solutions of the standard constraint equations on an asymptotically Euclidean space-like hypersurface which are identical to Kerr initial data (or Schwarzschild initial data in the time-symmetric case) outside a compact set, see Fig.~\ref{fig:globalspacetime}. The result was proven by making use of the fact that the constraint equations are highly underdetermined and showing that this freedom in specifying solutions could be used to glue two solutions of the constraints together across an open region. 

The Corvino-Schoen theorem can be combined with the semi-global stability theorem mentioned above to prove the existence of a large class of space-times which are globally asymptotically flat. The argument proceeds roughly as follows (for details see~\cite{Chrusciel:2002}): on an asymptotically Euclidean space-like hypersurface $\Sigma$ construct initial data which are identical to Kerr
initial data outside some compact set. Use the standard Einstein evolution equations to evolve the initial data for some time into the future. In the domain of dependence of the Kerr initial data the space-time will be identical to the Kerr space-time, which is stationary. Hence, it can be conformally rescaled so that it admits a smooth conformal boundary. Pick a space-like hypersurface in the development of the non-Kerr initial data and extend it smoothly through the Kerr portion of the space-time so that in becomes asymptotically hyperboloidal. The evolved solution induces hyperboloidal data on this space-like hypersurface~$\Sigma^\uparrow$. If the original asymptotically Euclidean data on $\Sigma$ are small enough, then the hyperboloidal data on $\Sigma^\uparrow$ will be small enough so that the semi-global stability theorem implies the existence of a complete global future development of the initial data on~$\Sigma$. A similar argument can be made to obtain a global development of the same data on $\Sigma$ into the past so that in the end there exists a globally hyperbolic regular asymptotically flat space-time. This argument shows that there exists a large class of such space-times.

\begin{figure}[htb]
  \centering\begin{tikzpicture}[x=1cm,y=1cm]
  \def\pulse #1 width #2 color #3 scatter (#4) {
    \def\q{#1}
    \def\a{#2}
    \coordinate (c1) at (\q,\q-\X);
    \coordinate (c2) at ($(c1) + (\a,\a) $);
    \path let \p1=(c1) in coordinate (c3) at (0,{\x1+\y1});
    \path let \p1=(c3) in coordinate (c4) at (0,{\y1+2*\a})
                          coordinate (c6) at ({0.5*(\X-\y1)}, {0.5*(\X+\y1)})
                          coordinate (c7) at (\a, {\y1 + \a});
    \path let \p1=(c4) in coordinate (c5) at ({0.5*(\X-\y1)}, {0.5*(\X+\y1)});
    \fill[#3!30,opacity=0.5] (c1)--(c2)--(c4)--(c3)--cycle;
    \fill[#3!30,opacity=0.5] (c3)--(c4)--(c5)--(c6)--cycle;
    \draw[#3] (c1)--(c3)--(c6);
    \draw[#3] (c2)--(c4)--(c5);
    \foreach \t in {#4}
             \draw[#3!50]  ($ (c2)!\t!(c4) $) -- ++($ (i0)-(c2) $);
  }
  \def\X{4cm}
  \coordinate (ip) at (0, \X);
  \coordinate (im) at (0,-\X);
  \coordinate (i0) at (\X,0);

  \pulse 10mm width 3mm color green scatter () 
  \pulse 20mm width 4mm color magenta scatter () 
  \pulse 30mm width 2mm color blue scatter () 
  \pulse 37mm width 1mm color red scatter (0.1,0.12,...,0.2) 
  
  \draw[thick] (ip) node[above]{$i^+$} -- node[above right]{$\scri^+$} (i0) node[right]{$i^0$} -- node[below right]{$\scri^-$} (im) node[below]{$i^-$};
  \draw[thick,dashed] (im)  -- (ip) ;
  \draw (5mm,-\X+10mm) node{$\sM$};
  \draw[green!70] (0,0) -- (i0) node[below,pos=0.8,black]{$\Sigma$};
  \draw[fill=white] (ip) circle(2pt);
  \draw[fill=white] (i0) circle(2pt);
  \draw[fill=white] (im) circle(2pt);
\end{tikzpicture}
  \caption{Incoming and outgoing radiation pulses}
  \label{fig:inout}
\end{figure}
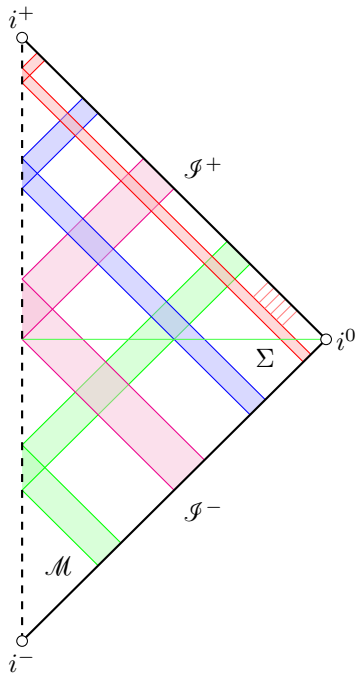

The construction just described, in particular the use of an exact solution near space-like infinity, seems awkward and unnecessary. However, there is a reason for this awkwardness which is the main focus of this paper and which we will try to clarify by referring to Fig.~\ref{fig:inout}
The figure shows the conformal diagram of an asymptotically flat space-time $\sM$ with its conformal boundary consisting of $\scri^\pm$, $i^\pm$ and $i^0$\footnote{This figure as well as the previous one are intended only as a symbolic illustration of the real situation. In both figures black holes are excluded. Furthermore, the propagation of the radiation does not take any tails into account.}. The null-infinities $\scri^\pm$ are null hypersurfaces. Each point on $\scri^+$ resp.\ $\scri^-$ corresponds to the limit point of a null geodesic in the infinite future resp.\ past. Similarly, the `points' correspond to limit points of time-like geodesics in the past and future and limit points of space-like geodesics. In general, the 'points' $i^\pm$ and $i^0$ do not exist as regular points in the embedding conformal manifold. A non-vanishing total ADM mass of the space-time creates a singularity near space-like infinity as can be seen most easily by examining the conformal extension of the Schwarzschild space-time. Furthermore, when matter is present, as it is in most physical circumstances, then the infinite time-like past and future will be singular unless the matter decays appropriately towards the future and emerges appropriately from the past. In this sense, the figure shows a prototypical example of a space-time with a complete null-infinity while, in general, an asymptotically flat space-time would only correspond to a part of this figure.

The figure also shows several pulses of radiation, entering the space-time at different (advanced) times. They all enter the physical space-time through $\scri^-$, cross the space-time, and go back out through $\scri^+$. The earlier a pulse enters $\scri^-$ the earlier it leaves it at $\scri^+$. The late ingoing radiation does not affect the earlier outgoing radiation after some time i.e., for every ingoing radiation pulse there is a time after which any outgoing pulse propagates unaffected towards $\scri^+$. The physical processes captured in this figure are the propagation of radiation through space-time and its scattering off a central body (indicated by the dashed line). Other physical processes which could happen are scattering off the background curvature of the space-time and scattering due to self-interaction of the radiation (such as for example a general Yang-Mills or gravitational radiation). These are shown in the figure only for the last ingoing pulse.

The aim of this diagram is to emphasize that there are many different \emph{physical} processes by which a point on $\scri^-$ can influence a point on $\scri^+$ via radiation propagating through $\sM$. However, the diagram suggests another causal connection between such points on $\scri^-$ and $\scri^+$, namely by a signal from the first point along $\scri^-$ towards $i^0$ and then from $i^0$ along $\scri^+$ to the second point. Ignoring for the moment the fact that, in a general asymptotically flat space-time, $i^0$ is not a regular `point', from the point of view of the conformal space-time in which $\sM$ is embedded this possibility is perfectly legitimate. However, from the physical point of view it is definitely illegitimate since it does not involve any piece of the physical space-time.

The existence of this possibility creates a problem for causality in the physical space-time because non-vanishing late ingoing radiation could influence the earliest outgoing radiation in a non-physical way. For this reason the unphysical causal relation between $\scri^-$ and $\scri^+$ is problematic and needs a thorough understanding. 

This phenomenon, which we may refer to as \emph{causality violation at infinity} (CVI) is only a problem when the radiation is described by equations which allow access to conformal infinity. Many field equations such as the Euler equations for ideal fluids or the massive wave equation cannot be extended in a regular to way to $\scri$ so that CVI does not occur. However, there are many important cases such as the Maxwell equations, the equations of gravitational perturbations or the conformally coupled wave equation where CVI could occur. In fact, we will see in the later sections of this paper indications based on analytical as well as numerical methods that these processes can and will occur unless they are explicitly suppressed.

Of course, there are many issues with the presented argument not the least of which is the fact that the figure suppresses two dimensions. While in Minkowski space-time space-like infinity is indeed a regular point, in general it is not regular and it may not even by a point. Interpreting the lines labeled $\scri^-$ resp.\ $\scri^+$ as single generators of the actual null hypersurfaces one issue is the behaviour of the incoming resp. outgoing radiation as they approach space-like infinity, i.e., in the infinite future resp. past along the generators. Another issue concerns the uniformity of this behaviour across generators. When space-like infinity is a regular point and the equations under consideration are conformally invariant, then $i^0$ is as good a point as any other and there is no question as to how information travels from $\scri^-$ to $\scri^+$ across $i^0$ as long as the fields are continuous. However, in the general case there are open questions: Under what conditions can information about the late data on $\scri^-$ travel illegitimately across $i^0$? What does this process do to this information and how does the emerging information at the other side on $\scri^+$ affect the legitimate physical information registered on $\scri^+$?

It is also possible to phrase this issue in terms of initial data imposed on the fields on an asymptotically Euclidean hypersurface $\Sigma$ as indicated in the figure. The incoming radiation induces data on $\Sigma$ which, in turn, give rise to outgoing radiation on $\scri^+$. The late time behaviour of the incoming radiation field is therefore intimately related to the asymptotic behaviour of initial data on $\Sigma$ at space-like infinity and therefore we can ask the question as to which conditions have to be imposed on initial data on an asymptotically Euclidean hypersurface in order to prevent the illegal unphysical propagation of information from $\scri^-$ to $\scri^+$ across $i^0$? 

Of course, the description of the issues just given is very superficial and lacks many details, part of which we will discuss in the following sections. One obvious simplification which was used above is that we consider the radiation as propagating on a given background. Clearly, this is not the case in the discussion of the general structure of asymptotically flat space-times where it is very difficult to separate the propagating degrees of freedom of gravitational waves from the underlying background. The terms `wave' and `background' do not even make sense in this context unless one provides a scale which in most circumstances is entirely arbitrary~\cite{Isaacson:1968,Isaacson:1968a}. This suggests to separate the problem into two parts. The first part is concerned with the propagation of fields on a given background while the second part involves the investigation of space-like infinity  for general asymptotically flat space-times. Even in the first part one needs to have an idea about the structure near space-like infinity but for a given space-time this is far easier than in the general case.

The structure of the paper will be as follows. In sec.~\ref{sec:structure-near-sl} we present the general discussion of the representation of space-like infinity of general asymptotically flat space-times based on the work of Friedrich. We present a brief discussion of the current knowledge in the general case. In sec.~\ref{sec:waves-spac-near} and~\ref{sec:numer-stud-infinity} we specialise to the Minkowski and Schwarzschild space-times and discuss analytical aspects, in particular the energy method, for fields in the neighbourhood of space-like infinity in sec.~\ref{sec:waves-spac-near}, while sec.~\ref{sec:numer-stud-infinity} is devoted to numerical studies of such fields.

The conventions used in this work are those from Penrose and Rindler~\cite{Penrose:1986}.

\section{The structure near space-like infinity}
\label{sec:structure-near-sl}

In this section we will briefly review what is known about the structure of space-times in the region near $i^0$. Referring to the figures~\ref{fig:globalspacetime} and \ref{fig:inout} it is obvious that this region must be highly special since --- from a physical point of view --- it is a region of infinite temporal and spatial extent while --- from the conformal point of view --- it fits into a finite region around the `point' $i^0$ and this statement is true for arbitrarily small neighbourhoods of $i^0$. The conformal compactification crams space-time events with vastly different separations close together. Clearly, when there are any differences in the behaviour of the gravitational field as one follows different directions to infinity then this will lead to problems of regularity when the infinite regions are squashed into a small neighbourhood of $i^0$. Unless the space-time is very regular there will be intrinsic singular behaviour near $i^0$ after the compactification. In fact, we know that in only very special cases such as for Minkowski space is space-like infinity a regular point. As soon as the space-time has a non-vanishing ADM mass there is no regular point $i^0$. This has been pointed out already by Penrose~\cite{Penrose:1965}. Even when we study fields other than the gravitational field on a space-time we can expect irregularities near space-like infinity even if the fields are otherwise regular on null-infinity.

However, the compactification process is not completely arbitrary; it preserves the conformal structure of the space-time. Making use of this property is the key to disentangle the properties of the region near $i^0$ and to get more detailed information about the behaviour of the gravitational and other fields there. In our presentation we will follow the original work by Friedrich~\cite{Friedrich:1998,Friedrich:2003,Friedrich:1995,Friedrich:2003a}, delegating more detailed arguments and proofs to these references. 

\subsection{Asymptotically flat space-times}
\label{sec:asympt-flat-space}

Let us start by setting the arena. We define a \emph{conformal space-time} as a triple $(\sM,g_{ab},\CF)$ for which the following three conditions are satisfied: (i) $(\sM,g_{ab})$ is a (time and space orientable) Lorentz manifold, (ii) $\CF$ is a smooth scalar field on $\sM$ so that the set $\tilde{\sM} = \{p\in \sM: \CF(p)>0 \}$ is non-empty and connected, and (iii)~the \emph{gravitational field} $K_{abc}{}^d:=\CF^{-1}C_{abc}{}^d$, i.e., the rescaled Weyl tensor, extends smoothly to all of $\sM$. We call two conformal space-times $(\sM,g_{ab},\CF)$ and $(\hat{\sM},\hat{g}_{ab},\hat{\CF})$ equivalent if $\sM$ and $\hat\sM$ are diffeomorphic and (after identifying $\sM$ and $\hat\sM$ with a suitable diffeomorphism) there exists a smooth strictly positive function $\theta:\sM\to \RR$ such that $\hat{g}_{ab}=\theta^2 g_{ab}$ and $\hat\CF = \theta \CF$.

These definitions imply that $(\tilde\sM,\tilde g_{ab})$  is an open submanifold of $\sM$ with a Lorentzian metric $\tilde g_{ab}:=\CF^2g_{ab}$ which is defined independently of which representative of a class of conformal manifolds is used. We assume for simplicity that $\tilde g_{ab}$ satisfies the vacuum Einstein equations (with vanishing cosmological constant) on $\tilde\sM$ (even though this is only necessary in some neighbourhood of $\scri$). Under these conditions we call $(\tilde\sM, \tilde g_{ab})$ the \emph{physical space-time} and $(\sM,g_{ab})$ the \emph{unphysical space-time}. By construction, the physical space-time allows the attachment of a conformal boundary as outlined by Penrose~\cite{Penrose:1986,Penrose:1964a,Penrose:1963}. The boundary is given by $\scri:=\{p\in \sM:\CF(p)=0,\quad \dd\CF(p) \ne 0 \}$, a regular hypersurface in $\sM$. While our definitions admit more general cases we will restrict ourselves to the case where $\scri$ has two connected components $\scri^+$ and $\scri^-$ each with the topology $S^2\times \RR$. This is the physically most interesting case and it is the only possible case when the assumption is made that every null geodesic in $\sM$ intersects each of $\scri^-$ and $\scri^+$ exactly once or never (see the usual definition of asymptotically flat space-times).
Furthermore by (iii), the Weyl tensor $C_{abc}{}^d$ associated with $g_{ab}$ (which is identical to the Weyl tensors of $\hat g_{ab}$ and $\tilde g_{ab}$) vanishes on $\scri$.

\subsection{The conformal field equations}
\label{sec:conf-field-equat}

The geometry of $(\sM,g_{ab},\CF)$ is described by a number of geometric quantities related by a hierarchy of equations. The metric $g_{ab}$ gives rise to a torsion-free connection, its Levi-Civita connection, i.e., a covariant derivative operation $\nabla_a$ for which the metric is covariantly constant, so that
\begin{equation}
  \label{eq:1}
  \nabla_c g_{ab}=0
\end{equation}
holds.

The Riemann tensor $R_{abc}{}^d$ of $\nabla_a$ can be split into the conformally invariant Weyl tensor $C_{abc}{}^d$ and the Schouten tensor $P_{ab}$
\begin{equation}
  \label{eq:2}
  R_{abc}{}^d = C_{abc}{}^d - P_{ac}\delta_{b}^d + P_{a}{}^dg_{bc} + P_{bc}\delta_{a}^d - P_b{}^dg_{ac} .
\end{equation}
The Schouten tensor is determined by the Ricci tensor $R_{ab}$ and vice versa by the equations
\[
  R_{ab} = -2 P_{ab} - P_c{}^c g_{ab} \iff P_{ab} = -\frac12 \left( R_{ab} - \frac16 R g_{ab}\right).
\]
Finally, the Riemann tensor satisfies the Bianchi identity
\begin{equation}
  \label{eq:3}
  \nabla_{[e} R_{ab]c}{}^d = 0.
\end{equation}
This generic geometric setting is specialised by imposing the physical condition in the form of the vacuum Einstein equations \emph{on the physical metric}. In order to do this we need to relate the physical fields with the unphysical fields. Let $\tnabla_a$, $\tC_{abc}{}^d$ and $\tP_{ab}$ be the connection of the physical metric $\tg_{ab}$ and its related Weyl and Schouten tensors. With $\Upsilon_a = \nabla_a(\log\CF)$ the conformal transformations between the physical and unphysical fields hold on $\tilde \sM$ (here, $\alpha_a$ is any covector field on $\tilde\sM$)
\begin{align}
  \label{eq:4}
    \tnabla_a \alpha_b &= \nabla_a \alpha_b + \Upsilon_a\alpha_b + \Upsilon_b \alpha_a - g_{ab} \Upsilon^e\alpha_e,\\
    \tC_{abc}{}^d &= C_{abc}{}^d,\label{eq:5}\\
    \tP_{ab} &= P_{ab} + \nabla_a \Upsilon_b + \Upsilon_a \Upsilon_b - \frac12 g_{ab} \Upsilon^c\Upsilon_c.\label{eq:6}
\end{align}
In view of \eqref{eq:4} and for later use we associate with every covector $\lambda_a$ a $(1,2)$-tensor defined by $[\lambda]^a_{bc}: = \delta^a_b\lambda_c + \lambda_b\delta^a_c - g_{bc}g^{ae}\lambda_e$.

The physical Riemann tensor also satisfies the second Bianchi identity $\tnabla_{[e}\tR_{ab]c}{}^d=0$. In view of its decomposition into physical Weyl and Schouten tensors similar to~\eqref{eq:2} and the vacuum Einstein equations $\tR_{ab}=0=\tP_{ab}$ this identity implies the Bianchi equation for the Weyl tensor
\begin{equation}
  \label{eq:7}
  \tnabla_{[e} C_{ab]c}{}^d = 0 \iff \tnabla_e C_{abc}{}^e = 0.
\end{equation}
The conformal transformation of the connection together with the conformal invariance of the Weyl tensor implies the equation for the gravitational field $K_{abc}{}^d=\CF^{-1}C_{abc}{}^d$
\begin{equation}
  \label{eq:8}
  \nabla_e K_{abc}{}^e = 0.
\end{equation}
This equation is at the core of the theory of gravitation. It describes the propagation of the degrees of freedom carried by gravitational waves. The equation looks superficially like the spin-2 equation for a zero-rest-mass field, i.e., a higher spin generalisation of the free Maxwell or Weyl equations. In the present vacuum case the equation has no source terms. In more general circumstances the gravitational waves would be driven by matter terms given by the Ricci (and, therefore, by the energy-momentum) tensor.

We can now list the hierarchy of equations that determine an unphysical space-time such that the related physical space-time satisfies the vacuum Einstein equations. In order to do this we need to introduce coordinates $(x^\mu)_{\mu=0:3}$ and a tetrad field $(e_\ka)_{\ka=0:3}$ on $\sM$. Let $\gamma^\kc_{\ka\kb}$ denote the connection coefficients defined with respect to the tetrad field by $\nabla_{e_{\ka}}e_\kb = \gamma^\kc_{\ka\kb} e_\kc$. We assume the tetrad vectors to form an orthonormal basis so that $g(e_\ka,e_\kb) = \eta_{\ka\kb} = \mathrm{diag}(+1,-1,-1,-1)$. Under these conditions the connection coefficients satisfy $\gamma_{\ka\kc\kb}:=\eta_{\kc\ke}\gamma^\ke_{\ka\kb}=-\gamma_{\ka\kb\kc}$. The tetrad vectors have components with respect to the coordinates defined by the expansion $e_\ka = e_\ka^\mu\del_\mu$.

Now we can write down the equations for the frame components and the connection coefficients
\begin{align}
  \label{eq:9}
  e_\ka(e_\kb^\mu) &-   e_\kb(e_\ka^\mu) - \gamma_{\ka\kb}^\kc e_\kc^\mu +  \gamma_{\kb\ka}^\kc e_\kc^\mu=0,\\
  e_\ka(\gamma_{\kb\kc}^\kd) &- e_\kb(\gamma_{\ka\kc}^\kd) - \gamma_{\ka\kc}^\ke \gamma_{\kb\ke}^\kd + \gamma_{\kb\kc}^\ke \gamma_{\ka\ke}^\kd - \gamma_{\ka\kb}^\ke \gamma_{\ke\kc}^\kd + \gamma_{\kb\ka}^\ke \gamma_{\ke\kc}^\kd\label{eq:10}\\ &\hspace{5em}= \CF K_{\ka\kb\kc}{}^\kd - P_{\ka\kc}\delta_{\kb}^\kd + P_{\ka}{}^\kd \eta_{\kb\kc} + P_{\kb\kc}\delta_{\ka}^\kd - P_\kb{}^\kd \eta_{\ka\kc}.\nonumber
\end{align}
In contrast to these `structural' equations the remaining equations are tensorial and can be written down without reference to a particular coordinate system and tetrad:
\begin{align}
  \nabla_a \nabla_b \CF + \CF P_{ab}  - g_{ab} S &=0,  \label{eq:11}\\
  \nabla_a S + P_a{}^c\nabla_c\CF  &= 0,\label{eq:12}\\
  \nabla_a P_{bc} - \nabla_b P_{ac} - \nabla_e\CF K_{abc}{}^e &=0, \label{eq:13}\\
  \nabla_e K_{abc}{}^e &=0\label{eq:14}.
\end{align}
The equation~\eqref{eq:11} is recognised as imposing the vanishing of the trace-free part of  $\tP_{ab}$ in \eqref{eq:6}, introducing the variable $S$ for the trace. Equation~\eqref{eq:13} is a consequence of the Bianchi identity \eqref{eq:3} expressed in terms of $K_{abc}{}^d$ and using the Bianchi equation~\eqref{eq:8}. Equation~\eqref{eq:12} for the trace term $S$ is a consequence of the other equations and closes the system. It may seem as if these equations do not impose the vanishing of the physical scalar curvature $\tilde\Lambda$. However, \eqref{eq:6} implies $\CF S - \frac12 \nabla_e\CF\nabla^e\CF = 6 \tilde\Lambda$ and a brief calculation using \eqref{eq:11} and \eqref{eq:12} shows that the left hand side is constant on $\sM$, so that $\tilde\Lambda=0$ if it vanishes at some point in~$\sM$.

Let us make a couple of remarks about the structure of the system (\ref{eq:9}-\ref{eq:14}) of conformal field equations. Choosing $\CF=1$ shows that the equations (\ref{eq:11}-\ref{eq:14}) reduce to imposing the vanishing of the trace-free part of $P_{ab}=\tP_{ab}$ and, as a consequence, the constancy of its trace. The other equations yield different parts of the vacuum Bianchi identities. Eq.~\eqref{eq:10} is the definition of the Riemann tensor components in terms of the connection coefficients. Taking an appropriate trace yields, together with \eqref{eq:9}, a system of equations which is equivalent to the standard Einstein equations in first-order form.

The system can be understood as giving equations for the set of unknowns $u:=(e_\ka^\mu,\gamma_{\ka\kb}^\kc,P_{ab}, \linebreak[1]K_{abc}{}^d,\CF,S)$. It is invariant under three kinds of gauge transformations which are due to the way the representation of the physical metric is chosen: these are the choice of a coordinate system, the choice of a tetrad and, in addition, the choice of the conformal factor $\CF$, i.e., the way the conformal class of $\tilde g_{ab}$ is represented. In order to get a well-posed system of equations these choices have to be fixed and we will discuss a particularly advantageous choice in due course. 

As in the case of the standard Einstein equations one needs to split the equations into evolution and constraint equations. This can be done in complete analogy to the standard case by a $(1+3)$ decomposition of $\sM$. We will not present the details here but refer instead to~\cite{Friedrich:1983a,Friedrich:2002,Frauendiener:2004a} and references therein. The result is analogous to the standard Einstein equations. The full system can be split into two sets, one of which contains all equations  intrinsic to the leaves $\Sigma_t$ of a space-like foliation of $\sM$ into level-sets of a global time coordinate $t$. The other set contains equations for the change of the geometric quantities transverse to the leaves. The first set is referred to as the constraint equations while the second set contains the evolution equations. One finds the usual property that `the constraints are propagated by the evolution'. This means that solutions of the constraint equations at one instant of time $t$ are mapped by the evolution into solutions of the constraint equations at a later time $t'$. This implies that it is enough to prepare initial data for the evolution equations on an initial hypersurface $\Sigma_0$ by solving the constraint equations, then the solution of the evolution equations will produce a solution of the full set.

The concrete character of the constraint and evolution equations is not fixed. For example, in the standard formulation of the Einstein equations one can pick a choice of coordinates and particular combinations of unknowns to rewrite the constraint equations into equations for them which are elliptic (this is the most studied case), or parabolic~\cite{Bartnik:2004}. Most recently, it was shown that the constraints even admit a hyperbolic formulation~\cite{Racz:2015,Racz:2016,Beyer:2017a,Beyer:2019}. To date, the study of the conformal constraints has not progressed as far as for the standard equations (see however \cite{Butscher:2002,Butscher:2007}). This is mainly due to the fact that one can construct solutions to the constraints of the conformal field equations from solutions of the standard constraints~\cite{Andersson:1992,Andersson:1993}.

In a similar way, one can obtain different types of evolution systems by choosing appropriate variables and gauges. The most useful formulations are those which provide strongly or symmetric hyperbolic equations. A lot of work has been devoted to the discussion of different ways to formulate the evolution equations. There are no preferred formulations in the sense that every formulation has its particular drawbacks and advantages. For more details on this particular aspect of the evolution equations we refer to the literature (see e.g., \cite{Alcubierre:2012} and \cite{Baumgarte:2010})

For the conformal field equations similar considerations apply. In order to obtain equations for which a well-posed initial (boundary) value problem (I(B)VP) can be formulated judicious choices of variables and gauges have to be made. Fortunately, in the context of the conformal representation of space-times there exists a very powerful gauge, the conformal Gauß gauge, which we will introduce next.

\subsection{The conformal Gauß gauge}
\label{sec:conformal-gau3-gauge}

In order to discuss this gauge we need to step back and discuss a more general class of connections related to a conformal class of metrics. Let $g_{ab}$ be a metric on $\sM$ and let $[g_{ab}]$ denote its conformal class, i.e., the set of all metrics $\hat{g}_{ab}$ for which there exists a function $\theta:\sM \to \RR^+$ so that $\hat{g}_{ab}=\theta^2 g_{ab}$. A \emph{Weyl connection} is a torsion-free connection which is compatible with the conformal class $[g_{ab}]$ in the sense that there exists a 1-form $\lambda_a$ on $\sM$ such that
\[
  \nabla_c g_{ab} = -2 \lambda_c g_{ab}.
\]
Every torsion-free connection $\tnabla_a$ which is compatible to a metric $\tg_{ab}=\theta^2g_{ab}$ in the conformal class is a Weyl connection since
\[
  \tnabla_c g_{ab} = -2 \theta^{-1}\nabla_c \theta g_{ab} \implies \lambda_a = \theta^{-1}\nabla_a\theta = \nabla_a \log\theta.
\]
The converse is not true in general since not every 1-form on $\sM$ is exact.

Any two Weyl connections $\nabla_a$ and $\tnabla_a$ are related by the usual formula: let $\omega_a=\tilde\lambda_a - \lambda_a$ be the difference of the two corresponding Weyl forms, then
\begin{equation}
\label{eq:15}
\tnabla_a v_b = \nabla_a v_b - \omega_a v_b - \omega_b v_a + \omega^c v_c g_{ab} = \nabla_a v_b - [\omega]_{ab}^c v_c. 
\end{equation}
The curvature tensors, i.e., the Weyl and Schouten tensors, of these two Weyl connections are related as follows
\begin{align}
  \label{eq:16}
    \tilde{C}_{abc}{}^d &= C_{abc}{}^d,\\
    \tilde{P}_{bc} &= P_{ab} - \nabla_a \omega_b + \omega_a \omega_b - g_{ab}\omega^c \omega_c.\label{eq:17}
\end{align}
These transformations are almost identical to equations~(\ref{eq:4}--\ref{eq:6}). They coincide when $\tnabla_a$ is the connection compatible with $\tg_{ab}$ and $\nabla_a$ is compatible with $g_{ab}=\CF^2\tg_{ab}$ since then $\omega_a = -\Upsilon_a$. In contrast to a connection which is compatible to a metric the Schouten tensor for a general Weyl connection is no longer symmetric, which is clear from the appearance of the term $\nabla_a\omega_b$ which is symmetric only for closed 1-forms $\omega_b$.

Before we discuss the consequences of introducing Weyl connections we need to give a generalisation of the notion of geodesics into the conformal setting. Consider a space-time $(\tilde\sM,\tg_{ab})$ with compatible connection $\tnabla_a$. A \emph{conformal geodesic} is a curve $\gamma$ in $\tilde\sM$ with a tangent vector $u^a$ together with a 1-form $h_a$ defined along $\gamma$ such that the equations
\begin{align}
  \label{eq:18}
  u^a\tnabla_a u^b + 2 (u^c h_c)u^b - (u^cu_c)h^b &=0,\\
  u^a\tnabla_ah_b - (u^ch_c)h_b + \frac12 (h^ch_c)u_b + u^a\tilde{P}_{ab} &=0\label{eq:19}
\end{align}
hold along the curve. The curve $\gamma$ is determined by the conformal class of $\tg_{ab}$. This can be easily checked by replacing $\tg_{ab}$ with any metric in the conformal class and $\tnabla_a$ with an arbitrary Weyl connection. Using the transformation formulae for connection and Schouten tensor one finds that the equations are invariant. Given a starting point and initial vector $u^a$ and 1-form $h_b$ these equations uniquely determine the curve $\gamma$ (and hence its tangent vector $u^a$) and a 1-form $h_a$ defined on it.

In view of the fact that \eqref{eq:18} can be written as $u^a\tnabla_au^b+[h]^b_{ac}u^au^c=0$ we  define the new Weyl connection $\hnabla_a$ using $h_b$ as a Weyl form and find that the equations for a conformal geodesic reduce to
\begin{align}
  \label{eq:20}
  u^a\hnabla_a u^b &=0,\\
 u^a\hP_{ab} &=0\label{eq:21}.
\end{align}
This shows that the curve $\gamma$ is a geodesic for this new connection which is special in the sense that its Schouten tensor is aligned with the geodesic.

We are now in the position to introduce the conformal Gauß gauge. Let us start with the physical space-time $(\tilde\sM,\tg_{ab})$ and a space-like hypersurface $\tSigma$ in $\tilde\sM$. We assume that $\tg_{ab}$ satisfies the vacuum Einstein equations $\tilde{R}_{ab}=0 \iff \tilde{P}_{ab}=0$. Let $p\in \tSigma$ be an arbitrary point and let $\underline{u}^a$ be the unit-normal at $p$ to $\tSigma$. At $p$ select a 1-form $\underline{h}_a$ then there exists a conformal geodesic $\gamma$ with parameter $\tau$ through $p=\gamma(0)$ with tangent vector $u^a$ and a 1-form $h_a$ which agree with $\underline{u}^a$ and $\underline{h}_a$ at $p$. Furthermore, selecting a coordinate system $(x^1,x^2,x^3)$ of $\tSigma$ we can obtain a coordinate system $(x^0=\tau,x^1,x^2,x^3)$ for $\tilde\sM$ near $p$ by dragging the coordinates along the conformal geodesics. In a similar way we can obtain a basis of orthogonal vectors $e_\ka$ near $\tSigma$ by selecting an orthonormal triad $(e_1,e_2,e_3)$ of tangent vectors to $\tSigma$ at $p$, defining $e^a_0=u^a$ and using parallel transport with $\hnabla_a$ to define a tetrad at every point in a neighbourhood of $\tSigma$. Thus, each of the basis vectors $e^b_\kc$ satisfies the equation
\begin{equation}
  \label{eq:22}
  u^a\tnabla_ae^b_\kc = - h_au^a e^b_\kc - h_ae^a_\kc u^b + u_ae^a_\kc h^b.
\end{equation}
Note, that $u^a$ is parallel along $\gamma$ due to \eqref{eq:20}. Also note, that the vectors $e_\ka$ will be mutually orthogonal but, in general, will not be unit-vectors with respect to the metric $\tg_{ab}$ since $\hnabla_a$ is compatible with the conformal class $[\tg_{ab}]$ but not necessarily with the metric itself. This means that there is a function $\CF$ defined by $\tg_{\ka\kb} = \tg(e_\ka,e_\kb) = \CF^{-2} \eta_{\ka\kb}$ and that $(e_\ka)_{\ka=0:3}$ will be an orthonormal tetrad with respect to the metric $g_{ab}$ defined by $g_{ab}=\CF^2 \tg_{ab}$.

This shows that by using the congruence of time-like conformal geodesics emanating from a space-like hypersurface $\tSigma$ we can define
\begin{enumerate*}[label=(\roman*)]
\item a system of adapted coordinates,
\item a basis of orthogonal vectors $e_\ka$,
\item a particular Weyl connection defined by the Weyl form $h_a$, and
\item a particular representative $g_{ab}$ of the conformal class of the vacuum metric $\tg_{ab}$ defined by the conformal factor $\CF$.
\end{enumerate*}
This collection of specific choices is referred to as the \emph{conformal Gauß gauge}.

One can say more about the conformal factor. Since $\CF^{-2}=\tg_{ab}u^au^b$ we can use \eqref{eq:18} and \eqref{eq:19} to derive an equation for the behaviour of $\CF$ along any conformal geodesic. Using the fact that $\tilde{P}_{ab}=0$ we find the equation
\[
  \dddot\CF = 0
\]
which implies that $\CF(\tau)$ is a quadratic function of $\tau$ with coefficients determined by the initial values\footnote{Here, and in what follows, we regard underlined quantities as constant along the conformal geodesics.} $\underline{u}^a$ and $\underline{h}_a$ given as functions of the coordinates $(x^1,x^2,x^3)$ on $\tSigma$. Defining $\underline{\CF}=\CF(0)$, $\underline{h}_0=\underline{h}_a\underline{u}^a$ and $\underline{H}=\underline{h}_a\underline{h}_b\tg^{ab}$ we find
\begin{equation}
  \label{eq:23}
  \CF(\tau) = \underline{\CF} \left(\frac14 \frac{\underline{H}}{\underline{\CF}^2}\tau^2 + \underline{h}_0 \tau + 1 \right).
\end{equation}
Thus, this conformal factor is \emph{completely known} as a function of the adapted coordinates, and we find that it has at least one zero unless $\underline{h}_a=0$, in which case the conformal geodesic becomes a metric geodesic for the physical metric $\tg_{ab}$.

A similar result holds for the components $b_\ka=b_ae^a_\ka$  of the 1-form $b_a = \CF h_a$ with respect to the parallel basis $(e_\ka)_{\ka=0:3}$. Using the equations \eqref{eq:19} and \eqref{eq:22} and $u^a=e^a_0$ one can derive equations for the components $(b_0,b_1,b_2,b_3)$
\begin{equation}
  \label{eq:24}
  \dot b_0 = - \frac1{2\CF} \left(b_0^2 - b_1^2 - b_2^2 - b_3^2  \right), \qquad \dot b_k = 0 ,
\end{equation}
with the explicit solutions
\begin{equation}
  \label{eq:25}
  b_0 =  \frac{\underline{b}_0^2 - \underline{b}_1^2 - \underline{b}_2^2 - \underline{b}_3^2}{2\underline{\CF}}\tau + \underline{b}_0, \qquad
  b_k = \underline{b}_k.
\end{equation}

\subsection{The reduced conformal field equations}
\label{sec:reduc-conf-field}

Let us recall the conditions that characterise the conformal Gauß gauge:
\begin{enumerate}
\item the coordinates are dragged along $u^a=e_0^a$: $e^\mu_0=\delta^\mu_0$ or $u^a=\del_\tau^a$,
\item the frame $(e_\ka)$ is parallel with respect to $\hnabla_a$: $\hgamma^\kc_{0\kb}=0$,
\item the Schouten tensor $\hP_{ab}$ is aligned with $u^a$: $\hP_{0\ka}=0$,
\item the conformal factor $\CF$ is given explicitly by \eqref{eq:23}.
\end{enumerate}
The reduced conformal field equations are then obtained by writing (\ref{eq:9}), (\ref{eq:10}) and (\ref{eq:13}) with respect to the Weyl connection defined by the gauge and extracting those equations which contain a time derivative, i.e., a derivative in the direction of $e_0$. This yields the equations
\begin{align}
  \label{eq:26}
    \del_\tau e_\kb^\mu  &=-  \hgamma_{\kb 0}^\kc e_\kc^\mu,\\
  \del_\tau\hgamma_{\kb\kc}^\kd  &=- \hgamma_{\kb 0}^\ke \hgamma_{\ke\kc}^\kd + \CF K_{0\kb\kc}{}^\kd + \hP_{\kb\kc}\delta_{0}^\kd - \hP_\kb{}^\kd\eta_{0\kc} + \hP_{\kb 0}\delta_{\kc}^\kd ,\label{eq:27}\\
  \del_\tau\hP_{\kb\kc} &= \hgamma_{\kb 0}^\ka \hP_{\ka\kc} + b_\ke K_{0\kb\kc}{}^\ke. \label{eq:28}
\end{align}
These equations have to be completed with \eqref{eq:14} for the rescaled Weyl tensor. We note that $\CF$ and $b_\ke$ are both explicitly given as functions of the coordinates by~\eqref{eq:25} once the initial data for the conformal geodesics have been fixed. Therefore, \eqref{eq:11} and \eqref{eq:12}, which provide equations for the conformal factor $\CF$, reduce to identities in the conformal Gauß gauge. 

It is remarkable that (\ref{eq:26}--\ref{eq:28}) are \emph{ordinary differential equations} along the conformal geodesics. This has the consequence that the information about the propagation of the gravitational waves is entirely confined to \eqref{eq:14}. This is a tremendous simplification compared to other gauges, and it is the key to getting a grip on the region near space-like infinity.

\subsection{The cylinder at space-like infinity}
\label{sec:cylinder-at-sl}

Let us now turn our attention to space-like infinity. In order to capture the essence of this notion in the context of the CFE we need to impose conditions which we glean from the prototype of flat Minkowski space and its conformally compactified representation as a submanifold with boundary embedded inside the Einstein cylinder~\cite{Penrose:1963}. In that picture, a space-like  hypersurface $\tSigma$ which extends to $i^0$ becomes a cross-section $\Sigma$ of the Einstein cylinder, a topological 3-sphere obtained from $\tSigma$ by suitable attachment of a point at infinity. We assume that this holds in general, i.e., that $\Sigma$ is a manifold obtained from $\tSigma$ by adding a point $i$. The metrics $\tildeh_{ab}$ on $\tSigma$ and $h_{ab}$ on $\Sigma$ are supposed to be conformally related via a conformal factor $\cf$ defined on $\Sigma$ which is smooth and strictly positive on $\tSigma=\Sigma \backslash \{i\}$ and satisfies $\cf(i)=\dd \cf(i)=0$ and $(\mathrm{Hess}\,\cf)_{ab} = -2 h_{ab}$. In contrast to the flat case $\cf$ cannot be smooth at $i$ unless the ADM mass of the space-time vanishes.

For the next step we introduce normal coordinates centred at $i$. Given a unit-vector $\ub \in T_{i}\Sigma\sim S^2$ we denote by $\gamma_{\ub}$ the geodesic starting at $i$ in the direction $\ub$. Then, with every $\rho>0$, there is a point $P(\rho,\ub)=\gamma_{\ub}(\rho)=\gamma_{\rho\ub}(1)$. Restricting the range of $\rho$ to a suitable open interval $(0,a)$ for some $a>0$ we can assume that there is a bijection between $(0,a)\times S^2$ and the `punctured ball' $B^*$ of radius $a$ around $i$, (i.e. the solid ball with the origin $i$ removed).

Let $\pi:T_BS^2\to B$ be the sphere bundle over $B$ and consider the map $j:(-a,a)\times S^2 \to T_BS^2$ given by $j(\rho,\ub)= \dot{\gamma}_{\ub}(\rho)$. This is a smooth embedding of $(-a,a)\times S^2$ into  $T_BS^2$ with $\pi \circ j(\rho,\ub) = \gamma_{\ub}(\rho)$. Since $\gamma_{\ub}(\rho)= \gamma_{\rho\ub}(1)= \gamma_{-\ub}(-\rho)$ every point in $B$ has two pre-images except for $i$ whose pre-image is an entire sphere. The punctured ball $B^*$ is diffeomorphic to $(0,a)\times S^2$ and to $(-a,0)\times S^2$.

Setting $\hat{B}=\mathrm{im}(j)$ and lifting the metric $h_{ab}$ via $\pi$ up to $\hat{B}$ the net-effect of the previous construction is the replacement of the point $i$ with a sphere $I^0$ which smoothly connects $B^*$ with another copy of $B^*$, i.e., we have blown up the point $i$ to a sphere and extended the manifold $\Sigma$ beyond $I^0$ by attaching a collar. In what follows we will always think of spatial infinity in terms of the sphere $I^0$ and not in terms of the point $i$. 

Given suitable initial data on $\tSigma$ and hence on $\Sigma$ we can construct the congruence of time-like conformal geodesics based on $\Sigma$. These initial data are chosen so that $b_a=\nabla_a\CF$ with the additional condition that $\dot\CF=0$ on $\Sigma$. We also specify the initial conformal factor $\underline{\CF}$ to be $\cf/\kappa$ where $\kappa$ is a strictly positive function on $\tSigma$, which we will specify later. The coordinates $(\rho,\ub)$ on (a neighbourhood of $I^0$ in) $\Sigma$ are propagated along the congruence onto an open set in $\sM$ where they are complemented by $\tau$ to form a 4-dimensional coordinate system.

The conformal factor $\CF$ and 1-form $b_a$ are given explicitly by
\begin{equation}
  \label{eq:29}
  \CF(\tau) = \frac{\cf}{\kappa} \left(1-\frac{\kappa^2}{\omega^2}\tau^2\right), \qquad b_0(\tau) = - 2 \Omega \frac{\kappa^2}{\omega^2}\tau,\qquad b_i(\tau) = e_i(\cf).
\end{equation}
The function $\omega$ introduced here is defined in terms of $\cf$ by
\begin{equation}
  \label{eq:30}
  \omega^2 = -\frac{4\cf^2}{h^{ab}\del_a\cf\del_b\cf}.
\end{equation}
Eq.~\eqref{eq:29} describes the behaviour of the conformal factor $\CF$ along the conformal geodesics starting at a given point $q$ on $\Sigma$. With the initial conditions specified the conformal factor vanishes at $\tau_\pm(q)=\pm\tfrac{\kappa(q)}{\omega(q)}$, indicating that the conformal geodesic intersects $\scri^\pm$ at a time which is specified by the initial conditions at $q$. Thus, the location of $\scri^\pm$ is known \emph{a priori}. From the conditions imposed on the conformal factor $\cf$ at $i$ one can deduce that $\omega = \rho + O(\rho^2)$ near $I^0$. Thus, if $\kappa$ is non-zero on $I^0$ then $\scri^\pm$ are reached for $\tau_\pm=0$, i.e. they occur at the point $i$ (or, after blowup, on the sphere $I^0$). This is familiar from the example of Minkowski space-time embedded in the Einstein cylinder where $\scri^\pm$ can be regarded as the future/past light-cone of the single point $i^0$.

However, there are other possibilities: if $\kappa=\rho \mu$ for a strictly positive function $\mu$ on $\Sigma$ with $\mu(i)=\mu_0$, then $\tau_\pm(i) = \mu_0$ so that the intersections with $\scri^\pm$ occur at the times $\tau_\pm=\pm\mu(i)$.

\begin{figure}
    \centering
    \includegraphics[width=0.7\textwidth]{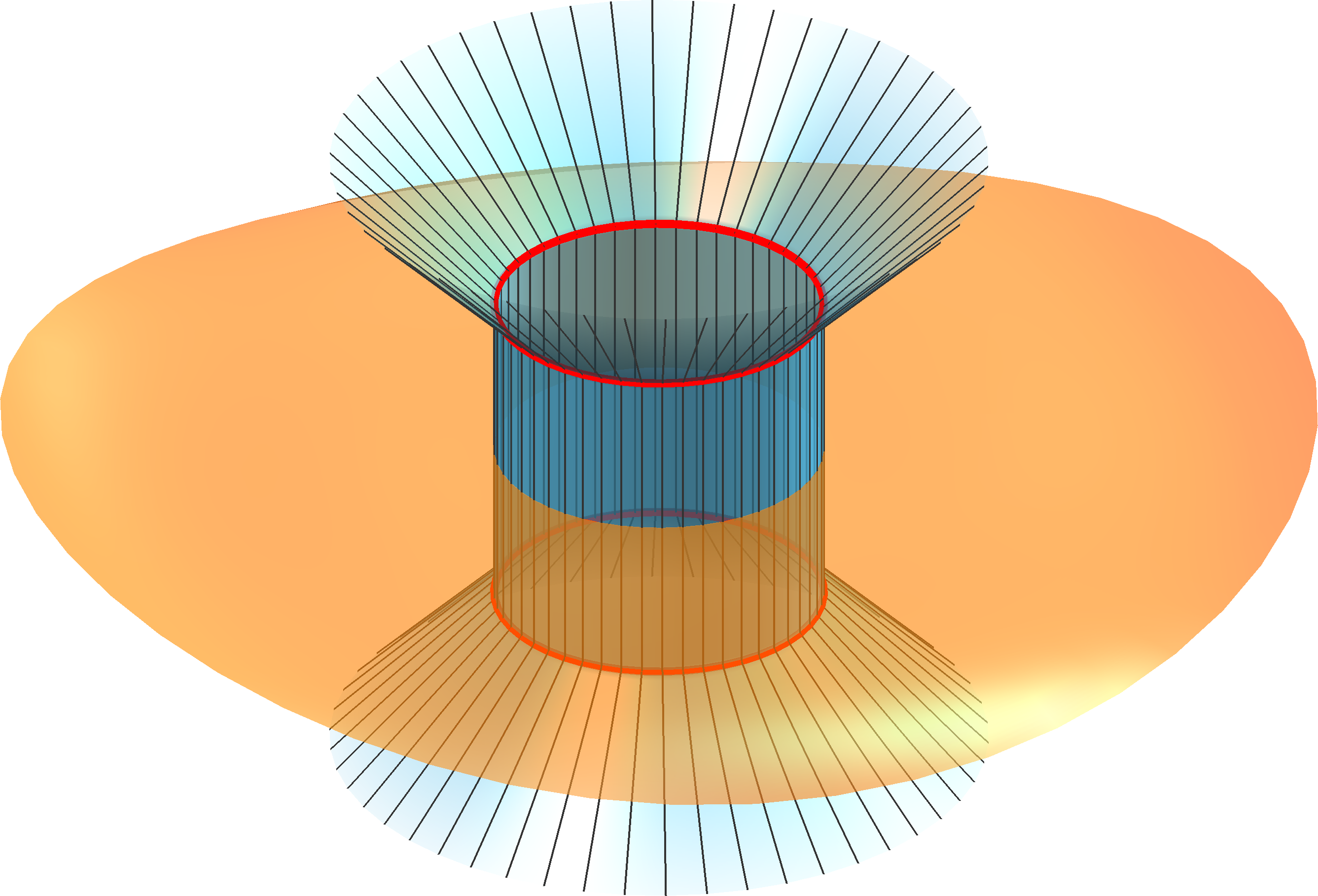}
    \caption{The cylinder at space-like infinity together with the space-like hypersurface $\Sigma$, the null-infinities $\scri^\pm$ and the critical sets $I^\pm$ (red). One dimension is suppressed.}
    \label{fig:cylinder}
\end{figure}

These observations motivate the introduction of the following sets: the null-infinities $\scri^\pm=\{p \in \sM: \tau = \pm\tfrac{\kappa}{\omega}, \quad \rho > 0\}$, the initial sphere $I^0=\{p \in \sM: \tau=0,\quad \rho=0\}$, the \emph{cylinder at space-like infinity} $I=\{p\in \sM: \rho = 0, \quad -\mu_0 < \tau < \mu_0\}$ and the \emph{connecting spheres} $I^\pm=\{p\in \sM : \rho=0, \quad \tau=\pm\mu_0\}$. These are particularly interesting since they are the locations where the cylinder $I$ and the null-infinities $\scri^\pm$ meet. Fig.~\ref{fig:cylinder} shows a rendering of the geometry at space-like infinity.

\subsection{Critical sets and the behaviour of fields at infinity}
\label{sec:sing-sets-behav}

The previous sections showed how the full power of conformal geometry can be used to set up an arena which is suitable for the discussion of fields near space-like and null infinity. Clearly, not all fields will be amenable to this procedure. Since we represent a physical space-time in terms of a conformally equivalent geometry, the fields need to be compatible with the conformal structure, i.e., they have to behave appropriately under conformal rescalings of the metric.

In the later parts of this work we will discuss in detail certain `test fields' on a given fixed space-time which is represented by a manifold which includes the cylinder at space-like infinity. For the remainder of this chapter we will give a very short overview of the more challenging case where we consider the CFE in order to determine a vacuum space-time which admits the construction of the cylinder at space-like infinity.

To this end we assume that we have such a space-time and restrict ourselves to a neighbourhood of space-like and null infinity. Specifically, we assume that we are given a suitable neighbourhood of the point $i$ in a space-like hypersurface $\Sigma$ with a conformal factor $\cf$ satisfying the conditions given above. We think of $i$ as blown up to the sphere $I^0$ and the neighbourhood as being extended beyond $I^0$. We construct the normal coordinates $(\rho,\ub)$ and extend them via the congruence of conformal geodesics to a 4-dimensional neighbourhood covered by the coordinates $(\tau,\rho,u,v)$ with any two coordinates $(u,v)$ covering the sphere having chosen the initial data for the congruence as given above with a function $\kappa=\rho$ (for simplicity) which exhibits the cylinder $I$ as the set given by $\rho=0$ and $-1<\tau<1$. Note, that the conformal factor $\CF$ vanishes for $\rho=0$, i.e., on $I^0$ and since it is dragged along the conformal geodesics it also vanishes on $I$ which is generated by those curves. We specify a frame $(e_1,e_2,e_3)$ which is orthonormal with respect to the metric $h_{ab}$ on $\Sigma$ in such a way that $e_1$ is aligned along the radial direction emanating out of $I^0$. Then the frame vectors $e_2$ and $e_3$ are tangential to the spheres of constant $\rho$ on $\Sigma$. There remains a rotational degree freedom around $e_1$. This frame is well-defined in a neighbourhood of $I^0$ for positive and negative values of $\rho$ and extends via parallel transport along the conformal geodesics into a 4-dimensional neighbourhood of~$I^0$.

In this setting the conformal field equations are (\ref{eq:26}--\ref{eq:28}) together with~\eqref{eq:14}. As we have seen in sec.~\ref{sec:reduc-conf-field} the first three equations are ODEs along the conformal geodesics for the frame components $e^\mu_\kb$, the connection coefficients $\gamma^\ka_{\kb\kc}$ and the Schouten tensor $\widehat{P}_{ab}$. The equations contain the explicit expressions for the conformal factor $\CF$, the 1-form components $b_\ka$ and the remaining unknown $K_{abc}{}^d$ of the system. The rescaled Weyl tensor $K_{abc}{}^d$ in turn is governed by the `spin-2 equation' \eqref{eq:14}
\[
  \nabla_e K_{abc}{}^e = 0.
\]
For simplicity we expressed it using the connection compatible with $g_{ab}$ instead of the Weyl connection $\hnabla_a$. This equation is formally analogous to Maxwell's equations for the electromagnetic field tensor $F_{ab}$, and it shares many properties with it. The gravitational field tensor $K_{abc}{}^d$ can be split into two symmetric trace-free tensors $E_{ab}$ and $B_{ab}$, which are spatial in the sense that $u^aE_{ab} = u^aB_{ab} = 0$. Combined into one quantity $\psi_{ab}=E_{ab} + \mathrm{i} B_{ab}$ they contain five complex degrees of freedom, which are conveniently expressed in terms of complex combinations of components with respect to the frame $(e_\ka)_{\ka=0:3}$. For the details of this analysis we refer to~\cite{Friedrich:1998} for the general case and to sec.~\ref{sec:waves-spac-near} for the linearised equation.

The spin-2 equation can be split into five complex evolution equations for the components $\psib:=(\psi_i)_{i=0:4}$ together with three constraint equations. It turns out that the evolution equations can be written in the rather symbolic form
\begin{equation}
  \label{eq:31}
A^\tau  \del_\tau \psib + A^\rho \del_\rho \psib + A^u \del_{u} \psib + A^v \del_v \psib = S(\gamma,e)\psib  .
\end{equation}
Here, $S$ is a $5\times5$-matrix valued function depending on the other system variables $(e_\ka^\mu,\gamma^\ka_{\kb\kc})$, while the $5\times5$-matrices $A^\tau,\ldots, A^v$ depend only on the frame components $(e_\ka^\mu)$, and all matrices are Hermitian. This system extends smoothly to $I^0$ and beyond. It turns out that the system is symmetric hyperbolic in a neighbourhood of $\Sigma$. Thus, given smooth initial data there exists a smooth solution for some time into the future and the past. Of course these initial data must satisfy the constraint equations. We will not discuss this aspect of the problem in much detail (see, however, \Sectionref{sec:energyspin2}) because there are many important developments but refer to \cite{Friedrich:2004,Butscher:2002,Friedrich:1998,Andersson:1992,Daszuta:2019,Corvino:2006,Corvino:2000,Chrusciel:2003,Acena:2011,Dain:2001}.

The matrices $A^\tau$ and $A^\rho$ have specific properties which are fundamental for the behaviour of the solutions near the cylinder. First, one finds that the matrix $A^\rho$ vanishes for $\rho=0$, i.e., on the cylinder itself. This means that the system does not contain any outward derivative on $I$, it is entirely intrinsic to $I$. The cylinder becomes a total characteristic of the system. It completely separates the physical part of the space-time outside of $I$ from the unphysical part with $\rho<0$ in that no information can travel from one part to the other. Furthermore, the behaviour of solutions on the cylinder $I$ is completely determined by their initial conditions on $I^0$. 

Since the system reduces to an intrinsic system on $I$, the \emph{transport equations}, its evolution is determined by the matrix $A^\tau$ which is diagonal on $I$ with $A^\tau=\mathrm{diag} (1+\tau,1,1,1,1-\tau)$. Thus, it is positive definite everywhere except for the spheres $I^\pm$. The transport system is symmetric hyperbolic on $I$ except for the spheres $I^\pm$ where the hyperbolicity breaks down. This indicates a singularity of the transport equations and we refer to the spheres $I^\pm$ as the \emph{critical sets} for the transport equations.

The transport equations connect the sphere $I^-$ where $\scri^-$ touches the cylinder with $I^+$ where $\scri^+$ is attached. Thus, they connect the late incoming radiation with the early outgoing radiation via $I$. The criticality of the sets $I^\pm$ have the consequence that the solutions may develop singularities as they evolve along $I$ which manifest themselves on $I^+$. However, since the system \eqref{eq:31} is hyperbolic in a neighbourhood of $I$ except for the critical sets it is very likely that the singularities in $I^+$ will be carried away by the evolution along $\scri^+$, thereby spoiling the genuine physical information about the outgoing radiation. This would constitute the causality violation at infinity discussed in the introduction. It is an unphysical influence on the outgoing radiation caused by late ingoing radiation and generated by the unphysical evolution along the cylinder.

Therefore, in order to prevent the CVI, it is important to pin down the conditions under which the generation of a singularity on $I^+$ may occur. In the general case of the Einstein equations this problem is very complicated and not finally resolved. For this reason we focus in the remainder of the article on test fields on given backgrounds because they share some of the general properties but they are much more amenable to controlled numerical and analytical treatments.

We close this section with a few observations about the general case (see \cite{Friedrich:2013b,Kroon:2016}). Most of what is known refers to the case where the initial data on $\Sigma$ are time-symmetric i.e., for which the extrinsic curvature vanishes, and for our discussion we will assume that this is the case. The analysis of the transport system along $I$ reveals that, generically, the solution contains poly-logarithmic terms of the form $c (1-\tau)^k\log^j(1-\tau)$ for some positive integers $k$ and $j$. If $k$ is sufficiently small and if these singularities persist during the evolution along $\scri^+$, then this would have the consequence that Sachs' peeling property~\cite{Sachs:1960} and hence the smoothness of $\scri^+$ would be lost. Analogous statements apply for $\scri^-$. By taking derivatives in the $\rho$ direction one can derive transport equations for all the derivatives of all the system variables of any order $p$ in $\rho$, collectively denoted as $u^p$. Hence, in principle, one can control the evolution of a formal asymptotic expansion of the fields from initial data on $I^0$ along the cylinder. Similar statements to the one above also apply to the coefficients $u^p$: in general they will contain poly-logarithmic terms as above and they will become singular on the critical sets $I^\pm$.

The appearance of the poly-logarithmic terms can be suppressed in each order $p$ by imposing conditions on the initial data of the fields on $I^0$. This in turn means that there are conditions on the asymptotic behaviour of Cauchy data on $\Sigma$. These `obstructions' to the smoothness of null-infinity fall into two categories. The first one is caused by the behaviour of the conformal class of the metric $h_{ab}$ on $\Sigma$. It turns out that one part of the conditions on the initial data amounts to the vanishing of the (conformally invariant) Cotton-York tensor $Y_{ab}$ and the symmetric trace-free parts of all its derivatives at $i$. It is known that these conditions are satisfied for initial data which behave near $i$ like conformally compactified asymptotically flat static data. The second cause for the poly-logarithmic terms is due to the conformal scaling of the metric $h_{ab}$. The analysis of Kroon~\cite{Kroon:2012,ValienteKroon:2010} shows that poly-logarithmic terms will be present unless the initial data agree at $i$ with conformally rescaled static data.

It appears that the notion of asymptotic staticity is crucial here. In fact, Cauchy data which are conformal to static data in a neighbourhood of $i$ satisfy all the conditions to all orders so that for these data there are no obstructions and the solutions propagate smoothly up to $I^+$ and from there to $\scri^+$ without a loss of smoothness. This relates back to the discussion of the Corvino (and, more generally, the Corvino-Schoen) result on initial data, which are identical to Schwarzschild data near $i$. In that case, the solutions are stationary near space-like infinity and no issues of the kind discussed here arise with the evolution. However, the analysis seems to suggest that it may not be necessary that the conformal staticity holds in a full neighbourhood of $i$. Instead, it looks like it is sufficient for smoothness that the asymptotic expansions of the Cauchy data and a conformally static space-time agree at $i$ to all orders. It is, however, not clear whether this requirement is necessary. In fact, the clarification of this question is the main outstanding problem in this area at the moment (see~\cite{Friedrich:2013b,Friedrich:2018} for the latest developments).

Lifting the restriction of time-symmetry of the initial data will presumably have the consequence that the asymptotics would have to be that of a stationary space-time instead of a static one. It is unclear to what extent the asymptotic conditions on the expansion coefficients of the fields carry physical information. Superficially, they relate to the late-time behaviour of ingoing radiation and the early-time behaviour of outgoing radiation. However, there is no conceivable physical process that could put these two types of radiation into relation with each other apart from back-scattering. But this process could not be maintained asymptotically towards the critical sets. Therefore, it seems reasonable to conjecture that there is no physical information in the obstructions so that, conversely, \emph{no physically meaningful information is lost} by imposing the conditions formulated above at $i$. 

While the general resolution of this question is still outstanding, there has been some work on the linear case, i.e., the spin-2 equation on Minkowski space as well as other types of conformally invariant systems propagating in the asymptotic regions of more general backgrounds such as the Schwarzschild and the Kerr space-times. We will review these works in the following sections. Questions that are of interest in these cases are: Under what conditions can one prove existence and uniqueness for solutions? What are the generic smoothness properties of solutions? Are there similar conditions imposed on initial data near the cylinder $I$ at infinity to ensure a smooth propagation to the critical sets? Is there a physical meaning to these conditions? Do the singularities that occur on the critical sets propagate outwards to $\scri$ and if so, how is the smoothness affected? These and other questions have been studied both analytically and numerically as we will describe next.

\section{Waves on space-times near space-like and null infinity}
\label{sec:waves-spac-near}
\subsection{Motivation and background}
The asymptotics of wave-like fields on fixed background space-times is an arena of   very active research currently since the pioneering works in  \cite{christodoulou1986,klainerman1980,john1981,alinhac2003,tao2006,lindblad2008}. One of the motivations for these studies is their model problem character for more complicated scenarios, as for example the resolution of the black hole nonlinear stability problem for the full quasi-linear Einstein vacuum equations \cite{dafermos2019}. 

The cylinder representation of asymptotically flat globally hyperbolic space-times introduced in \Sectionref{sec:cylinder-at-sl} is a particularly attractive stage for analysing global properties of solutions of such equations because it provides simultaneous access to some of the most important asymptotic regimes: space-like and null infinity. 
To our knowledge, this has not been exploited in great detail in the literature yet, but see \cite{Friedrich:2003a,beyer2019a}.

One of the most important approaches to study the qualitative behaviour of solutions of hyperbolic (wave-type) partial differential equations (PDEs) is the \emph{energy method}. The purpose of this section is not to {introduce} the energy method. Moreover we assume that the reader is familiar with the basics of the Cauchy problem for hyperbolic equations; see \cite{john1971,bar2010,taylor2011} or \cite{Ringstrom:2009cj} and references therein. In this section we want to use  some of the results available in the literature to highlight certain particular conclusions that can be drawn from the energy method for fields on cylinder-type space-times, and what its limitations and problems are. We shall be especially interested in the asymptotics of the fields at the cylinder itself, at null infinity and at the \emph{critical sets} where these two regimes meet. To make life easy, we shall completely restrict to fields on the cylinder representation of Minkowski space-time (apart from some short discussions about the Schwarzschild space-time in \Sectionref{sec:nonlinearwaveenergy}). 
 The energy method is useful when exact solution techniques, numerical techniques, other approximation or other rigorous techniques are not available, are impractical or not desired for some other reason. Having said this, even the energy method can be very difficult in practice, and the information which can  be extracted from it is often very limited. 

As we will see the term `energy' is highly ambiguous. In fact, `mathematical energies' are in general not required to have any physical or geometric interpretation. While some geometric PDEs give rise to a \emph{continuity equation} which, if the background space-time has certain symmetries,
allows one to define \emph{conserved energies}, this may not always be the most useful choice of energy. 
The most useful energies are (positive) definite in some sense. Only in this case they give rise to some control of the unknown fields themselves. 
For so-called \emph{symmetric hyperbolic systems} (all the examples below can be reduced to symmetric hyperbolic form), there is always a positive definite energy and an associated estimate. In many cases, however, this estimate is too `weak' to allow for qualitative conclusions. In general, the energy method makes use of a remarkable
 interplay between the PDE and some underlying geometry with Stokes' theorem at its core. Even most `non-geometric energies' introduced for the purpose of the analysis have a geometric interpretation in terms of some auxiliary `artificial' geometry.

In most cases, the PDEs imply inequalities for the energies, so-called \emph{energy estimates}, which bound certain spatial integrals of the unknown variables, and, if possible, a number of their derivatives as well. If such energy estimates control `sufficiently many' derivatives, Sobolev embedding results \cite{taylor2011,adams1975} yield \emph{pointwise} estimates for the unknowns and some typically smaller number of derivatives. Our discussion here should not be understood as an introduction to any such delicate mathematical concepts. 

In this whole discussion we consider solutions of the \emph{Cauchy problem} (initial value problem) of a PDE with \emph{Cauchy data} imposed on some space-like hypersurface. Given certain assumptions, solutions of \emph{linear} PDEs are guaranteed to exist, with some fixed regularity, globally in time and space (unless the space-time or the equation `breaks down' in some way). The energy method for linear PDEs is therefore interesting to draw conclusions about the \emph{asymptotic behaviour} -- in the cylinder picture this means the behaviour at the boundaries, in particular at null infinity and the critical sets.
For \emph{nonlinear} equations however the situation is far more complicated. Solutions of the Cauchy problem may break down after a finite time well before any boundaries of the underlying space-time, and may therefore not exist long enough to even ask the question about their asymptotics. The first step for nonlinear PDEs is therefore always to study the \emph{global-in-time existence problem} \cite{taylor2011,christodoulou1986, klainerman1980, lindblad2008} by combining \emph{local-in time} energy estimates (like in the linear case) with \emph{continuation} and \emph{bootstrap arguments}, typically (but not exclusively) under \emph{small data restrictions}. Energy estimates therefore play a double role for nonlinear equations: (i) to establish global-in-time existence, and, (ii) to estimate the qualitative asymptotic behaviour of the solutions.

\subsection{The spin-2 system}
\label{sec:energyspin2}
After this very general discussion we continue now with the concrete problem of the \emph{spin-2 system}
\begin{equation}
  \label{eq:sp2spinorial}
  \nabla^A{}_{A'} \phi_{ABCD} = 0.
\end{equation}
The field $\phi_{ABCD}$ is a totally symmetric spinor field with five complex degrees of freedom closely related to the quantity $\psib$ in \Eqref{eq:31}. To simplify our discussions here we refer to \cite{Penrose:1986} for all the details about spinors.
 Many of the results presented in this subsection are building on \cite{Friedrich:2003a}.
The spin-2 system can be understood as a model for linearised gravitational fields on fixed backgrounds, in particular flat Minkowski space, or as a subsystem of the full nonlinear Einstein’s vacuum equations. We shall exploit its conformal invariance and study it as a field on the fixed space-time metric
\renewcommand*{\d}{\mathrm{d}}
\begin{equation}
 g 
   = \d t^2+\frac{2t}{r}\,\d t\,\d r-\frac{1-t^2}{r^2}\,\d r^2
     -\d\omega^2,
\end{equation}
where $\d\omega^2$ is the standard round unit metric on the $2$-sphere. In these coordinates, the cylinder representing space-like infinity is at $r=0$, and the time coordinate $t$ is restricted to the interval $(-1,1)$. The boundaries given by $t=\pm 1$ correspond to future/past null infinity.
More details in the derivation of this metric from the standard form of the Minkowski metric are given in \Sectionref{sec:spin2}. This metric can be understood as the cylinder representation of Minkowski space.
Once a spin-frame has been selected, \Eqref{eq:sp2spinorial} is equivalent to 
\begin{align} 
\label{0evol}
0 = \mathcal E_0 &= (1 + t)\,\partial_{t}\,\phi_0 -
r\,\partial_{r}\,\phi_0   -\eth\phi_1 + 2\,\phi_0,\\
0 = \mathcal E_1 &= 2\,\partial_{t}\,\phi_1
-\etp\,\phi_0 -\eth\phi_2 + 2\,\phi_1,   \\
0 = \mathcal E_2 &= 2\,\partial_{t}\,\phi_2
-\etp\,\phi_1 -\eth\phi_3,\\
0 = \mathcal E_3 &= 2\,\partial_{t}\,\phi_3
-\etp\,\phi_2 -\eth\phi_4 - 2\,\phi_3,\\
\label{4evol}
0 = \mathcal E_4 &= (1 - t)\,\partial_{t}\,\phi_4 
+ r\,\partial_{r}\,\phi_4  
-\etp\,\phi_3 - 2\,\phi_4, 
\end{align}
together with
 \begin{align}
\label{1con}
0 = \mathcal C_1 &= t\,\partial_{t}\,\phi_1 -
r\,\partial_{r}\,\phi_1 + \frac{1}{2}\etp\phi_0 -
\frac{1}{2}\eth\phi_2, \\
0 = \mathcal C_2 &= t\,\partial_{t}\,\phi_2 -
r\,\partial_{r}\,\phi_2 + \frac{1}{2}\etp\phi_1 -
\frac{1}{2}\eth\,\phi_3, \\
\label{3con}
0 = \mathcal C_3 &= t\,\partial_{t}\,\phi_3 -
r\,\partial_{r}\,\phi_3 + \frac{1}{2}\etp\phi_2 -
\frac{1}{2}\eth\,\phi_4. 
\end{align} 
We shall not introduce the operators $\eth$ and $\etp$ but instead refer again to \cite{Penrose:1986} for the details. For our discussion here it is sufficient to know that these are differential operators on the $2$-sphere.
The first group of equations is interpreted as \emph{evolution
  equations} and the second one as \emph{constraint
  equations}. It is evident that there are in principle infinitely many different `evolution equations' which can be extracted from \Eqref{eq:sp2spinorial}. For example one can always add multiples of \Eqsref{1con} -- \eqref{3con} to \Eqsref{0evol} -- \eqref{4evol}. The particular choice of evolution equations here is sometimes referred to as the \emph{boundary adapted system} \cite{Friedrich:1995} because it gives rise to a well-posed constraint-preserving initial {boundary} value treatment for the spin-2 system. While problems related to boundaries are important especially for numerical studies of the spin-2 system, see \Sectionref{sec:spin2}, no boundary-related issues occur  in this section due to our particular choice of boundaries below. 
It is interesting to observe that the `constraint equations' \Eqsref{1con} -- \eqref{3con} contain time-derivatives (and should therefore maybe not bear this name). The important point for us here is however that at $t=0$, where we shall agree to specify Cauchy data for the Cauchy problem in all of what follows, all time derivatives drop out there. We give a quick discussion of the Cauchy problem of the spin-2 system below.

For our treatment of the spin-2 field here we want to avoid most of the technicalities associated with spinor field geometry. For this reason we shall now consider expansions of $\phi=(\phi_0, \ldots, \phi_4)$ in terms of \emph{spin-weighted spherical harmonics} with corresponding \emph{spin weights} (again we refer to \cite{Penrose:1986} for the details). The complex expansion coefficients are labelled as $\phi^{lm}=(\phi^{lm}_0, \ldots, \phi^{lm}_4)$ for each integer $l\ge 0$ and $m=-l,\ldots,l$.   If $\phi$ is sufficiently smooth, knowing $\phi$ is equivalent to knowing the infinite family $\phi^{lm}$. Given an arbitrary solution $\phi\in C^1(\Omega)$, where $\Omega$ is some subset of the space-time manifold, the corresponding family $\phi^{lm}$, which only depends on $t$ and $r$,
satisfies the following infinite, but uncoupled, system of systems of PDEs (one for each $l=0,\ldots$ and $m=-l,\ldots,l$):
\begin{align} 
\label{0evolswsh}
0 = \mathcal E^{lm}_0 &= (1 + t)\,\partial_{t}\,\phi^{lm}_0 -
r\,\partial_{r}\,\phi^{lm}_0   +\alpha_2\phi^{lm}_1 + 2\,\phi^{lm}_0,\\
0 = \mathcal E^{lm}_1 &= 2\,\partial_{t}\,\phi^{lm}_1
-\alpha_2\phi^{lm}_0 +\alpha_0\phi^{lm}_2 + 2\,\phi^{lm}_1,   \\
0 = \mathcal E^{lm}_2 &= 2\,\partial_{t}\,\phi^{lm}_2
-\alpha_0\phi^{lm}_1 +\alpha_0\phi^{lm}_3,\\
0 = \mathcal E^{lm}_3 &= 2\,\partial_{t}\,\phi^{lm}_3
-\alpha_0\phi^{lm}_2 +\alpha_2\phi^{lm}_4 - 2\,\phi^{lm}_3,\\
\label{4evolswsh}
0 = \mathcal E^{lm}_4 &= (1 - t)\,\partial_{t}\,\phi^{lm}_4 
+ r\,\partial_{r}\,\phi^{lm}_4  
-\alpha_2\phi^{lm}_3 - 2\,\phi^{lm}_4,\\
\label{1conswsh}
0 = \mathcal C^{lm}_1 &= t\,\partial_{t}\,\phi^{lm}_1 -
r\,\partial_{r}\,\phi^{lm}_1 + \frac{1}{2}\alpha_2\phi^{lm}_0 +
\frac{1}{2}\alpha_0\phi^{lm}_2, \\
0 = \mathcal C^{lm}_2 &= t\,\partial_{t}\,\phi^{lm}_2 -
r\,\partial_{r}\,\phi^{lm}_2 + \frac{1}{2}\alpha_0\phi^{lm}_1 +
\frac{1}{2}\alpha_0\phi^{lm}_3, \\
\label{3conswsh}
0 = \mathcal C^{lm}_3 &= t\,\partial_{t}\,\phi^{lm}_3 -
r\,\partial_{r}\,\phi^{lm}_3 + \frac{1}{2}\alpha_0\phi^{lm}_2 +
\frac{1}{2}\alpha_2\phi^{lm}_4,
\end{align} 
where 
\[\alpha_0=\sqrt{l (l+1)}\quad\text{and}\quad
  \alpha_2=\begin{cases}
    \sqrt{l(l+1)-2}, & \text{for $l\ge 1$,}\\
    0, & \text{for $l=0$.}
  \end{cases}
\]  
For consistency we must assume that $\phi_0^{lm}=\phi_1^{lm}=\phi_3^{lm}=\phi_4^{lm}=0$ for $l=0$ and $\phi_0^{lm}=\phi_4^{lm}=0$ for $l=1$. It is easy to check that these assumptions are consistent with \Eqsref{0evolswsh} -- \eqref{3conswsh}. 


One needs to be careful because solutions of \Eqsref{0evolswsh} -- \eqref{3conswsh} (for all $l$ and $m$) might not always generate solutions of \Eqsref{0evol} -- \eqref{3con} (and therefore of \Eqref{eq:sp2spinorial}). However, under the following circumstances the two systems are equivalent. Let us focus on the evolution equations, and their Cauchy problem, first. Friedrich \cite{Friedrich:1986} was the first to realise that \Eqsref{0evol} -- \eqref{4evol} have a well-posed Cauchy problem. Given Cauchy data $\phi_*\in C^1(S\times\mathbb S^2)$ where $S\times\mathbb S^2$ is some subset of the, say, $t=0$-hypersurface, the Cauchy problem of \Eqsref{0evol} -- \eqref{4evol} has a unique solution $\phi\in C^1(D(S\times\mathbb S^2))$ where $D(S\times\mathbb S^2)$ is a subset of the space-time manifold known as the \emph{domain of dependence}; see below. Expanding the Cauchy data $\phi_*$ in terms of spin-weighted spherical harmonics as above, we get Cauchy data $\phi_*^{lm}\in C^1(S)$ for \Eqsref{0evolswsh} -- \eqref{4evolswsh} for each $l$ and $m$. It then turns out that the field $\phi$ generated from the family of uniquely determined solutions $\phi^{lm}\in C^1(D(S))$ of  the well-posed Cauchy problem of each system \Eqsref{0evolswsh} -- \eqref{4evolswsh} is identical to the solution $\phi$ of \Eqsref{0evol} -- \eqref{4evol} found before on $D(S\times S^2)$. The argument for this result makes essential use of the $C^1$-regularity requirement, i.e., it works for classical solutions, but not necessarily for weak solutions. In the following we exclusively analyse solutions $\phi^{lm}$ of \Eqsref{0evolswsh} -- \eqref{4evolswsh} generated by $C^1$-Cauchy data $\phi_*$ in this way. In fact, we shall without further notice simplify the assumption even further and assume that all solutions considered here are generated from Cauchy data $\phi_*$ in $C^\infty(S\times\mathbb S^2)$. This then allows us to take arbitrarily many derivatives of the solution everywhere on $D(S\times\mathbb S^2)$.

Consider now an arbitrary pair of integers $l$ and $m$ and a solution $\phi^{lm}$ of \Eqsref{0evolswsh} -- \eqref{4evolswsh} defined on $D(S)$ as above. The field $\phi^{lm}$ therefore determines the quantities $\mathcal C^{lm}_1$, $\mathcal C^{lm}_2$, $\mathcal C^{lm}_3$ defined in \Eqsref{1conswsh} -- \eqref{3conswsh} on $D(S)$. These fields might not vanish because we have not enforced the constraints yet. In any case, by differentiating the equations for $\mathcal C^{lm}_1$, $\mathcal C^{lm}_2$, $\mathcal C^{lm}_3$ with respect to $t$ and employing all other equations \Eqsref{0evolswsh} -- \eqref{3conswsh}, we can derive the so-called \emph{subsidiary system}:
\begin{align*}
\partial_{t}\mathcal C^{lm}_1 + \frac{1}{2}\alpha_0 \mathcal C^{lm}_2 + \mathcal C^{lm}_1 = 0,\\
\partial_{t}\mathcal C^{lm}_2 + \frac{1}{2}\alpha_0 \mathcal C^{lm}_3 - \frac{1}{2}\alpha_0 \mathcal C^{lm}_1 = 0,\\
\partial_{t}\mathcal C^{lm}_3 - \frac{1}{2}\alpha_0 \mathcal C^{lm}_2 - \mathcal C^{lm}_3 = 0.
\end{align*}
For each $l$ and $m$ this is a homogeneous linear system of ODEs in the `unknowns' $\mathcal C^{lm}_1$, $\mathcal C^{lm}_2$, $\mathcal C^{lm}_3$. This implies that $\mathcal C^{lm}_1$, $\mathcal C^{lm}_2$ and $\mathcal C^{lm}_3$ vanish identically (and therefore the constraint equations hold) on $D(S)$ if $\mathcal C_1^{lm}=\mathcal C_2^{lm}=\mathcal C_3^{lm}=0$ on $S$. Solving the Cauchy problem of \Eqsref{0evolswsh} -- \eqref{4evolswsh} for each $l$ and $m$ as above with Cauchy data picked to satisfy the constraints \Eqsref{1conswsh} -- \eqref{3conswsh} therefore yields a solution of the full (seemingly overdetermined) spin-2 system \Eqref{eq:sp2spinorial}.

We see easily that the evolution equations have the form
$A^{\mu}\,\partial_{\mu}\,\phi^{lm} = B\,\phi^{lm}$ and the matrix $A^{\mu}n_\mu$ defined for an arbitrary covector $n_\mu$ is Hermitian. Moreover, $A^{\mu}\nabla_\mu t = \mathrm{diag}(1 + t,\, 2,\, 2,\, 2,\, 1 - t)$. Hyperbolicity of the equations therefore only breaks down in the limit $t\rightarrow\pm 1$.
The domain of dependence $D(S)$ introduced above is bounded (independently of $l$ and $m$) by the characteristic surfaces of \Eqsref{0evolswsh} -- \eqref{4evolswsh}. The co-normal $n_\mu$ of such a surface is determined by the condition $\det A^\mu n_\nu=0$ \cite{john1971}. Here this yields the following characteristic \emph{curves} independently of $l$ and $m$:
\begin{equation}
  \label{eq:characteristiccurves}
  r(t)=r_*,\quad r(t)=\frac{r_*}{1-t},\quad r(t)=\frac{r_*}{1+t},
\end{equation}
for an arbitrary $r_*>0$. The first characteristic curve is `unphysical' in the sense that it is not invariant when we add multiples of the constraints to the evolution equations. The other two characteristics are however physical and coincide with null curves of $g$.
 This has the following immediate consequences. First, ${\scri}^{\pm}$ (given by the $t=\pm 1$-surfaces) are both characteristic and it is therefore not surprising that the spin-2 field may exhibit some singular behaviour in the limit $|t|\rightarrow 1$. Second, if the initial surface $S$ is of the form 
\[S=(r_-,r_+)\]
for arbitrary $0<r_-<r_+$, then the (future) domain of dependence is
\[D^+(S)=\left\{(r,t)\left|t\in\left[0,\frac{r_+-r_-}{r_++r_-}\right), r\in\left(\frac{r_-}{1-t},\frac{r_+}{1+t}\right)\right.\right\};\]
in all of what follows we focus on future evolutions from $t=0$ towards $\scri^+$ at $t=1$.

Since we are interested in the asymptotics of solutions $\phi^{lm}$ on $D^+(S)$ at $t=1$, we need to ask under which conditions the closure of $D^+(S)$ intersects $\scri^+$. This is clearly only the case if we take the limit $r_-\rightarrow 0$. To this end, we observe that \Eqsref{0evolswsh} -- \eqref{3conswsh} are perfectly well-defined on the cylinder (given by $r=0$). In fact the cylinder coincides with all characteristic curves \Eqref{eq:characteristiccurves} of \Eqsref{0evolswsh} -- \eqref{4evolswsh}. When we now include the cylinder in $S$, and therefore in $D^+(S)$, boundary conditions are necessary (or allowed) neither there at the \emph{inner boundary} $r=0$, nor at the \emph{outer boundary} at $r=r_+/(1+t)$. With this in mind, we define
\begin{align}
\label{eq:spin2setsFirst}
  D^+(r_*)&=\left\{(r,t)\left|t\in\left[0,1\right), r\in\left[0, \frac{r_*}{1+t}\right)\right.\right\},\\
  D_\tau^+(r_*)&=\left\{(r,t)\left|t\in[0,\tau], r\in\left[0, \frac{r_*}{1+t}\right)\right.\right\},\\
\label{eq:spin2setsLast}
S_\tau &=\left[0, \frac{r_*}{1+\tau}\right)\times \{\tau\},
\end{align}
for arbitrary $\tau\in[0,1)$. In all of what follows we fix an arbitrary $r_*>0$.

Let us now discuss the energy method for the spin-2 system on $D^+(r_*)$. Given an arbitrary sufficiently smooth solution $\phi$ of the evolution equations \Eqsref{0evol} -- \eqref{4evol} and an arbitrary $\tau\in [0,1)$, Friedrich \cite{Friedrich:2003a} showed that the identity
\begin{equation}
  \label{eq:firstenergyidentity}
  0=\int_{D_\tau^+(r_*)}\sum_{k = 0}^4 (\bar{\phi}^{lm}_k\,\mathcal E^{lm}_k + \phi^{lm}_k\,\bar{\mathcal E}^{lm}_k) dt dr,
\end{equation}
yields the \emph{energy estimate}
\begin{equation}
  \label{eq:firstenergyestimate}
E^{lm}(\tau)
\le (1 - \tau)^{-5}E^{lm}(0),
\end{equation}
where the \emph{energy} is defined as
\begin{equation}
  \label{eq:spin2firstenergy}
  E^{lm}(\tau)=\int_{S_\tau} (|\phi^{lm}_0|^2 + 2\,|\phi^{lm}_1|^2 + 2\,|\phi^{lm}_2|^2
+ 2\,|\phi^{lm}_3|^2 + |\phi^{lm}_4|^2)\,dr.
\end{equation}
It is a standard fact \cite{stein1970,sugiura1990} that this energy is equivalent to (the square of) the $L^2$-norm of $\phi$ over the $3$-dimensional space-like slices $S_\tau\times \mathbb S^2$ with the standard product metric on $\RR\times\mathbb S^2$, i.e., there are uniform constants $C_1,C_2>0$ such that
\[C_1\|\phi(\tau)\|^2_{L^2(S_\tau\times\mathbb S^2)} \le \sum_{l=0}^\infty\sum_{m=-l}^l E^{lm}(\tau)\le C_2\|\phi(\tau)\|^2_{L^2(S_\tau\times\mathbb S^2)}\]
for all $\tau\in [0,1)$. 

Before we continue let us develop some intuition of what this energy estimate means and elaborate on Friedrich's remark in \cite{Friedrich:2003a} that \Eqref{eq:firstenergyestimate} ``only gives little information on the behaviour of the solutions of the spin-2 system near $t = 1$''. 
Very roughly speaking, given that each component of $\phi$ appears quadratically in the energy in \Eqref{eq:spin2firstenergy}, the estimate \Eqref{eq:firstenergyestimate} appears to allow each of the fields $\phi_0$,\ldots, $\phi_4$ to blow up like $(1 - t)^{-5/2}$ in the limit $t\nearrow 1$. Is this the correct behaviour? To answer this question, we should not forget two things: Firstly, \Eqref{eq:firstenergyestimate} is just an \emph{estimate}. It might not be sharp in the sense that general solutions of the spin-2 system might behave `better' than $(1 - t)^{-5/2}$. Secondly, because \Eqref{eq:firstenergyidentity} does not impose the constraints, it is conceivable that the potentially singular behaviour described by \Eqref{eq:firstenergyestimate} is `unphysical', i.e., only solutions which violate the constraints exhibit this behaviour while solutions of the \emph{full} spin-2 system, evolution \emph{and} constraint equations together, possibly behave better.

In order to get a better picture, let us consider the following scalar model equation derived from `the most singular' part \Eqref{4evolswsh} of \Eqsref{0evolswsh} -- \eqref{4evolswsh}:
  \begin{equation}
    (1 - t)\,\partial_{t}u+ r\partial_{r}u - 2u=0.
  \end{equation}
This equation can be solved explicitly (e.g.\ using the method of characteristics), and we get
\begin{equation}
  \label{eq:scalarmodelexactsol}
  u(r,t)=\frac{u_*((1-t)r)}{(1-t)^2},
\end{equation}
for an arbitrary sufficiently smooth Cauchy data function $u_*(r)$. In general, this function therefore behaves like $(1-t)^{-2}$ close to $t=1$. If $u_*(r)=O(r)$ at $r=0$, it behaves like $(1-t)^{-1}$, and only if $u_*(r)=O(r^2)$, it is bounded at in the limit $t\nearrow 1$. This simple example exhibits one of the principles which we seem to rediscover in all of the following cases as well: The behaviour of the solutions \emph{at null infinity} is determined by their properties \emph{on the cylinder}. 
Now if we define an \emph{energy} in analogy to \Eqref{eq:spin2firstenergy} we find
\begin{align*}
  E(\tau)&=\int_{S_\tau} |u(r,\tau)|^2dr=\frac1{(1-\tau)^4} \int_{0}^{r_*/(1+\tau)} |u_*(r (1-\tau))|^2dr\\
&=\frac1{(1-\tau)^5} \int_{0}^{r_*\frac{1-\tau}{1+\tau}} |u_*(r)|^2dr,
\end{align*}
and therefore, in analogy to \Eqref{eq:firstenergyestimate}, obtain the \emph{energy estimate}
\begin{equation}
  \label{eq:scalarmodelestimate} 
  E(\tau)\le \frac1{(1-\tau)^5} E(0).
\end{equation}
We remark that we would obtain exactly the same energy estimate by following the above more abstract strategy used for the spin-2 evolution equations. The fact that this estimate follows from the explicit solution \Eqref{eq:scalarmodelexactsol} suggests that it is `as optimal as it can be' on the one hand.  On the other hand, however, this estimates is certainly not optimal because, firstly, $u$ seems to therefore behave like $(1-t)^{-5/2}$ (as opposed to $(1-t)^{-2}$ obtained from \Eqref{eq:scalarmodelexactsol}), and, secondly, the above calculation  makes it obvious that the estimate does  not take  \emph{information leaving the spatial domain through the outer boundary} into account correctly. Only in the limit $r_*\rightarrow\infty$ (provided the integrals exist and are finite), no information can leave the spatial domain and the estimate becomes optimal in the sense that the inequality sign in \Eqref{eq:scalarmodelestimate} can be replaced by an equal sign. In this case, the explicit solution \Eqref{eq:scalarmodelexactsol}  becoming `more and more spatially constant'  increases the divergence rate of the infinite spatial integral.

This last observation confirms another apparently general principle which we observe for the spin-2 system as well as the wave equation below: The more we differentiate the unknown with respect to $\rho$, the more this `spatially constant behaviour' is cancelled out and the more regular the energy integral at $\scri$ is therefore. This  was observed for the first time by Friedrich in \cite{Friedrich:2003a} and we shall now summarise and extend his arguments. Notice in particular that while \cite{Friedrich:2003a} performs the following analysis for solutions of the \emph{full} spin-2 system, we treat the  larger class of solutions of the spin-2 evolution equations (potentially constraint violating). Our results suggest that the phenomenology for both the full system and the subsystem comprised by the evolution equations is essentially the same. This result is relevant, for example, for numerical calculations for which the constraints can only be enforced initially and the numerical evolutions may therefore be driven into the constraint violating regime by numerical errors. It is also of interest for certain analytical arguments where solutions are constructed iteratively and the actual system of equations may only be satisfied in some limit. In all these cases it is  crucial  that constraint violating solutions behave essentially the same as constraint satisfying ones.

For each $l\ge 0$, $m=-l,\ldots,l$, let $\phi^{lm}$ be a sufficiently smooth vector of the components of the spin-2 field defined on $D^+(r_*)$ which solves \Eqsref{0evolswsh} -- \eqref{4evolswsh} as discussed above. At this stage we allow this solution to violate the constraints \Eqsref{1conswsh} -- \eqref{3conswsh}.
For arbitrary integers $p,q\ge 0$, the fields $\phi^{lm}_{k,(q,p)}=\partial_t^q\partial_r^p \phi_k^{lm}$ therefore satisfy
\begin{align} 
\label{0evolswshpq}
0 = \mathcal E^{lm}_{0, (q,p)} &= (1 + t)\,\partial_{t}\,\phi^{lm}_{0, (q,p)} -
r\,\partial_{r}\,\phi^{lm}_{0, (q,p)}   +\alpha_2\phi^{lm}_{1, (q,p)} - (p-q-2)\,\phi^{lm}_{0, (q,p)},\\
0 = \mathcal E^{lm}_{1, (q,p)} &= 2\,\partial_{t}\,\phi^{lm}_{1, (q,p)}
-\alpha_2\phi^{lm}_{0, (q,p)} +\alpha_0\phi^{lm}_{2, (q,p)} + 2\,\phi^{lm}_{1, (q,p)},   \\
0 = \mathcal E^{lm}_{2, (q,p)} &= 2\,\partial_{t}\,\phi^{lm}_{2, (q,p)}
-\alpha_0\phi^{lm}_{1, (q,p)} +\alpha_0\phi^{lm}_{3, (q,p)},\\
0 = \mathcal E^{lm}_{3, (q,p)} &= 2\,\partial_{t}\,\phi^{lm}_{3, (q,p)}
-\alpha_0\phi^{lm}_{2, (q,p)} +\alpha_2\phi^{lm}_{4, (q,p)} - 2\,\phi^{lm}_{3, (q,p)},\\
\label{4evolswshpq}
0 = \mathcal E^{lm}_{4, (q,p)} &= (1 - t)\,\partial_{t}\,\phi^{lm}_{4, (q,p)} 
+ r\,\partial_{r}\,\phi^{lm}_{4, (q,p)}  
-\alpha_2\phi^{lm}_{3, (q,p)} + (p-q-2)\,\phi^{lm}_{4, (q,p)}. 
\end{align}
In analogy to \Eqref{eq:firstenergyidentity} we then consider, for an arbitrary $l$ and $m$ as before, the identity
\begin{equation*}
  0=\sum_{k = 0}^4 \left(\bar{\phi}^{lm}_{k, (q,p)}\,\mathcal E^{lm}_{k, (q,p)} + \phi^{lm}_{k, (q,p)}\,\bar{\mathcal E}^{lm}_{k, (q,p)}\right),
\end{equation*}
which assumes the form 
\begin{equation}
\label{eq:spin2energyidpre}
\begin{split}
 0=&\partial_{t}\left((1 + t)|\phi^{lm}_{0, (q,p)}|^2+2|\phi^{lm}_{1, (q,p)}|^2+2|\phi^{lm}_{2, (q,p)}|^2+2|\phi^{lm}_{3, (q,p)}|^2+(1 - t)|\phi^{lm}_{4, (q,p)}|^2\right)\\
&-\partial_{r}\left(r|\,\phi^{lm}_{0, (q,p)}|^2- r|\,\phi^{lm}_{4, (q,p)}|^2\right)\\
&- 2(p-q-2) |\phi^{lm}_{0, (q,p)}|^2 + 4|\phi^{lm}_{1, (q,p)}|^2 - 4|\phi^{lm}_{3, (q,p)}|^2
+ 2(p-q-2)|\phi^{lm}_{4, (q,p)}|^2.
\end{split}
\end{equation}
This result does not look very useful. The trick \cite{Friedrich:2003a} is to decompose it as follows
\begin{equation}
\label{eq:spin2energyid}
\begin{split}
0=&\sum_{k=0}^3\Bigl[\partial_{t}\left((1 + t)|\phi^{lm}_{k, (q,p)}|^2\right)
+\partial_{t}\left( (1 - t)|\phi^{lm}_{k+1, (q,p)}|^2\right)\\
&- 2(p-q+k-2) |\phi^{lm}_{k, (q,p)}|^2 
+ 2(p-q-k+1)|\phi^{lm}_{k+1, (q,p)}|^2\Bigr]\\
&-\partial_{r}\left(r|\,\phi^{lm}_{0, (q,p)}|^2- r|\,\phi^{lm}_{4, (q,p)}|^2\right).
\end{split}
\end{equation}
So long as $p-q\ge 2$, it therefore follows that
\begin{align*}
  \partial_{t}&\sum_{k=0}^3\left((1 + t)|\phi^{lm}_{k, (q,p)}|^2\right)
\\
  \le &2(p-q+1)\sum_{k=0}^3(1+t)|\phi^{lm}_{k, (q,p)}|^2
+\partial_{r}\left(r|\,\phi^{lm}_{0, (q,p)}|^2- r|\,\phi^{lm}_{4, (q,p)}|^2\right).
\end{align*}
Integrating this inequality over $S_\tau$ for an arbitrary fixed $\tau\in [0,1)$ yields 
\begin{equation*}
  \partial_\tau E^{lm}_{0\ldots3, (q,p)}(\tau)\le 2(p-q+1) E^{lm}_{0\ldots3, (q,p)}(\tau)
\end{equation*}
for the \emph{energy}
\begin{equation*}
  E^{lm}_{0\ldots3, (q,p)}(\tau)=\int_{S_\tau}\sum_{k=0}^3\left((1 + t)|\phi^{lm}_{k, (q,p)}|^2\right)dr.
\end{equation*}
Integration in time (or, more precisely, the so-called Gr\"onwall lemma \cite{taylor2011,Ringstrom:2009cj}) yields the \emph{energy estimate}
\begin{equation}
  \label{eq:spin2estimatefull1N}
  E^{lm}_{0\ldots3, (q,p)}(\tau)\le E^{lm}_{0\ldots3, (q,p)}(0)e^{2(p-q+1)\tau}
\end{equation}
for all $\tau\in [0,1)$ provided $p-q\ge 2$. 

We notice that this energy estimate only controls the fields $\phi^{lm}_{0, (q,p)}$,\ldots, $\phi^{lm}_{3, (q,p)}$, and this only if $p-q\ge 2$. Observe that for $p-q<2$, the energy identity above does not separate the components $0,\ldots,3$ from component $4$ as nicely and the results become weaker. The case $p-q=0$ agrees identically with \Eqsref{eq:firstenergyestimate} and \eqref{eq:spin2firstenergy}. For $p-q=1$, \Eqref{eq:spin2energyidpre} yields nothing better than exactly the same estimate of the form \Eqref{eq:firstenergyestimate} as for $p-q=0$. In both these cases, the estimates encompass all $5$ components of $\phi$. The case $p-q<0$ is not of interest here. At this stage we have hence confirmed  that the solutions are more regular the more often we differentiate them with respect to $r$. 

What about $\phi^{lm}_{4, (q,p)}$? Supposing that $p-q\ge 2$ as before, the estimate 
\begin{align*}
  &\partial_{t}\left((1 + t)|\phi^{lm}_{0, (q,p)}|^2\right)
+\partial_{t}\left( (1 - t)|\phi^{lm}_{4, (q,p)}|^2\right)
+ 2(p-q-2)|\phi^{lm}_{4, (q,p)}|^2\\
&\le\sum_{k=0}^32(p-q+k-2) |\phi^{lm}_{k, (q,p)}|^2 
+\partial_{r}\left(r|\,\phi^{lm}_{0, (q,p)}|^2- r|\,\phi^{lm}_{4, (q,p)}|^2\right)
\end{align*}
obtained from
  \eqref{eq:spin2energyid} implies with the same arguments that
\begin{align*}
  &
\partial_{t}\int_{S_\tau}\left( (1 - t)|\phi^{lm}_{4, (q,p)}|^2\right) dr
+ 2(p-q-2) \int_{S_\tau}|\phi^{lm}_{4, (q,p)}|^2dr\\
&\le2(p-q+1) E^{lm}_{0\ldots3, (q,p)}(t) \le E^{lm}_{0\ldots3, (q,p)}(0)e^{2(p-q+1)t}\le E^{lm}_{0\ldots3, (q,p)}(0)e^{2(p-q+1)}
\end{align*}
where we have used \Eqref{eq:spin2estimatefull1N}. This means that
\begin{equation*}
 (1 - t)\partial_{t}E^{lm}_{4, (q,p)}
+ (2(p-q)-5) E^{lm}_{4, (q,p)}\le E^{lm}_{0\ldots3, (q,p)}(0)e^{2(p-q+1)}
\end{equation*}
with the \emph{energy}
\begin{equation*}
  E^{lm}_{4, (q,p)}(\tau)=\int_{S_\tau}|\phi^{lm}_{4, (q,p)}|^2 dr.
\end{equation*}
Integrating in time (or, more precisely employing Gr\"onwall's lemma) yields
\begin{align*}
E^{lm}_{4, (q,p)}(t)\le & (1-t)^{(2(p-q)-5)}E^{lm}_{4, (q,p)}(0)\\
&+\frac{1}{2(p-q)-5}\left(1- (1-t)^{(2(p-q)-5)}\right) E^{lm}_{0\ldots3, (q,p)}(0)e^{2(p-q+1)}.
\end{align*} 
This is bounded over $t\in [0,1)$ if $p-q\ge 3$, i.e., we expect $\phi_4$ to be slightly more singular at $t=1$ than $\phi_0,\ldots, \phi_3$.

How do we exploit this  information about energies? One common strategy is now to write the evolution equations formally as an ordinary differential equation interpreting especially all spatial derivative terms as source terms. Given that the energy estimates established before provide estimates for these source terms, this might allow us to solve the equation as an ordinary differential equation. As a consequence this might yield improved and more detailed control over the behaviour of the  unknowns. 

The following alternative less standard strategy is particularly well-suited to the cylinder representation of space-times and was first proposed by Friedrich 
in \cite{Friedrich:2003a} for the spin-2 system (we shall demonstrate below that the same approach also allows us to study the wave equation). The purpose is to split the spin-2 field up into potentially singular contributions from the cylinder and fully regular contributions from the `bulk space-time'.
Given the results about the energies above we can pick arbitrary integers $\rho\ge 0$, $q_0\ge 0$ and $p_0\ge 2+q_0+\rho$ and conclude from \Eqref{eq:spin2estimatefull1N} that
\begin{align*}
  \sum_{p+q=p_0+q_0}^{p_0+q_0+\rho}&\sum_{l=0}^\infty\sum_{m=-l}^l(1+l^2)^\rho\int_0^1 E^{lm}_{0\ldots3, (q,p)}(t)dt\\
&\le \frac{e^{2(p_0+\rho-q_0+1)}-1}{2(p_0-q_0-\rho+1)} \sum_{p+q=p_0+q_0}^{p_0+q_0+\rho}\sum_{l=0}^\infty\sum_{m=-l}^l(1+l^2)^\rho E^{lm}_{0\ldots3, (q,p)}(0),
\end{align*}
where the right hand side (and therefore the left hand side) is guaranteed to be finite as a consequence of our fundamental smoothness assumptions. Since the left hand side bounds the square of the Sobolev norm of order $\rho$ of the fields $\partial_t^{q_0}\partial_r^{p_0}\phi_0$, \ldots, $\partial_t^{q_0}\partial_r^{p_0}\phi_3$ over $D^+(r_*)\times\mathbb S^2$  \cite{adams1975,stein1970,sugiura1990}, it follows that
\[\partial_t^{q_0}\partial_r^{p_0}\phi_k\in H^\rho(D^+(r_*)\times\mathbb S^2)\]
for all $k=0,\ldots,3$ and arbitrary integers $\rho\ge 0$, $q_0\ge 0$ and $p_0\ge 2+q_0+\rho$. The Sobolev embedding theorem (see \cite{adams1975}, Theorem~4.12, Part~II) yields that therefore
\[\partial_t^{q_0}\partial_r^{p_0}\phi_k\in C^{\rho-3}\left(\overline{D^+(r_*)\times\mathbb S^2}\right)\] 
 if $\rho\ge 3$. Without going into any details of these sets, the essential point for us is that $\partial_t^{q_0}\partial_r^{p_0}\phi_0$,\ldots, $\partial_t^{q_0}\partial_r^{p_0}\phi_3$ and all of their partial derivatives of order $\rho-3$ therefore extend continuously to the boundary of $D^+(r_*)\times \mathbb S^2$, i.e., in particular to $\scri^+$ and the critical set at $t=1$. They are therefore \emph{regular} functions there. The same arguments applied to $\phi_4$ yield
for arbitrary $\rho\ge 3$, $q_0\ge 0$ and $p_0\ge 3+q_0+\rho$ that \[\partial_t^{q_0}\partial_r^{p_0}\phi_4\in C^{\rho-3}\left(\overline{D^+(r_*)\times\mathbb S^2}\right).\]
We conclude that  $\phi_4$ and all of its partial derivatives of order $\rho-3$ extend continuously to the boundary of $D^+(r_*) \times \mathbb S^2$.

All this now yields that
\[\partial_r^{5}\phi_0,\ldots, \partial_r^{5}\phi_3, \partial_r^{6}\phi_4\in C^{0}\left(\overline{D^+(r_*)\times\mathbb S^2}\right)\]
are nice regular functions with well-defined bounded limits at $t=1$.
Given the map $J$, which assigns to any uniformly continuous function $f$ on $D^+(r_*)\times\mathbb S^2$ the uniformly continuous function
\begin{equation}
  \label{eq:defJ}
  J[f](t,r,x)=\int_0^r f(t,\sigma,x) d\sigma,
\end{equation}
for all $(t,r,x)\in D^+(r_*)\times\mathbb S^2$, the identities
\[\phi_k(t,r,x)=\sum_{p=0}^4\frac{1}{p!}\partial_{r}^p\phi_k(t,0,x)r^p+J^5[\partial_r^{5}\phi_k](t,r,x)\]
for $k=0,\ldots,3$ and
\[\phi_4(t,r,x)=\sum_{p=0}^5\frac{1}{p!}\partial_{r}^p\phi_4(t,0,x)r^p+J^6[\partial_r^{6}\phi_4](t,r,x)\]
therefore cleanly separate the fields $\phi$ into  contributions from the cylinder at $r=0$ (the first sums; fully determined by intrinsic evolution equations on the cylinder, see \Eqsref{0evolswshpq} -- \eqref{4evolswshpq}) and contributions from the bulk of the space-time (the second terms). The bulk terms extend as regular functions to the boundary. Only the first terms can potentially be singular at $t=1$.
Hence \textbf{the question whether a solution $\phi$ of either the full spin-2-system, or just the spin-2-evolution equations, is singular or not at $\scri^+$ is determined on the cylinder}.
In the interest of brevity we refrain here now from discussing the evolution equations intrinsic to the cylinder and their potentially singular solutions at the critical set $t=1$.  We refer to \Sectionref{sec:spin2} and references there, in particular, \cite{ValienteKroon2002}.

As a last remark about the spin-2 system let us only mention in passing that the  spin-2 system has indeed a nice geometric energy defined by the so-called \emph{Bel-Robinson tensor} \cite{Szabados:2009ig}. In some cases, this energy is even conserved. We have not used this energy for any of our discussions here mainly because the conservation laws obtained for any $(p,q)$-derivative of the fields are in general not as natural as those for the fields themselves. In addition it should be emphasised that the Bel-Robinson energy is only natural when the \emph{full} spin-2 system is used. As explained before, this is however not always what we want.

\subsection{The conformally invariant scalar linear wave equation}
\label{sec:scalarwaveequationenergy}

We next discuss the conformally invariant scalar linear wave equation on the same cylinder representation of Minkowski space-time as in \Sectionref{sec:energyspin2}. This equation has been used as a numerical model problem, see in particular \Sectionref{sec:CWEMinkowski}. For us the main motivation
is to confirm and thereby strengthen the conclusions about wave-type equations on cylinder-type space-times found in \Sectionref{sec:energyspin2}.
 In addition the linear case here can also be seen as a preparation for the nonlinear case in \Sectionref{sec:nonlinearwaveenergy}. 

The equation of interest takes the form 
\begin{equation}
  \label{eq:waveenergy}
  (1-t^2)\del_{tt}f+2tr\del_{tr}f-r^2\del_{rr}f-2t\del_t f-\Delta_{\mathbb S^2} f=0
\end{equation}
where $\Delta_{\mathbb S^2}$ is the Laplace-Beltrami operator of the round unit metric on the $2$-sphere. One of major differences to the spin-2 system is that this is a single second-order PDE for a single unknown; there are therefore no issues regarding constraints and overdeterminateness. 

As for the spin-2 field, we proceed here by expanding the unknown $f$ with respect to (in this case scalar) spherical harmonics. This yields an infinite system (for each integer $l\ge 0$ and $m=-l,\ldots,l$) of scalar PDEs
\begin{equation}
  \label{eq:waveenergyswsh}
  (1-t^2)\del_{tt}f^{lm}+2tr\del_{tr}f^{lm}-r^2\del_{rr}f^{lm}-2t\del_t f^{lm}+l (l+1) f^{lm}=0,
\end{equation}
which is equivalent to \Eqref{eq:waveenergy} 
if we make analogous smoothness assumptions for the Cauchy problem as for the spin-2 system. It is easy to show that the quantity 
$f^{lm}_{(q,p)}=\partial_t^q\partial_r^p f^{lm}$ satisfies
\begin{equation}
\begin{split}
  (1-t^2)\del_{tt}f_{(q,p)}^{lm}&+2tr\del_{tr}f_{(q,p)}^{lm}-r^2\del_{rr}f_{(q,p)}^{lm}
+2(p-q-1)t\del_t f_{(q,p)}^{lm}\\
&-2(p-q)r\del_{r}f_{(q,p)}^{lm}
+(l (l+1)+C_{p,q}) f_{(q,p)}^{lm}
=0.
\end{split}
\end{equation}
with $C_{p,q}=(p-q)(1-(p-q))$ for arbitrary integers $p,q\ge 0$.

A straightforward (but lengthy) calculation reveals that the \emph{energy}
  \begin{equation}
    \label{eq:waveenergydef}
    \begin{split}
    E_{(q,p)}^{lm}(t)=\int_{S_t}\biggl[(1+l(l+1))\left|\sqrt{1-t}f_{(q,p)}^{lm}\right|^2
    &+(1+t)\left|(1-t)\partial_tf_{(q,p)}^{lm}\right|^2\\
    &+\left|\sqrt{1-t}\,r \partial_r f_{(q,p)}^{lm}\right|^2
    \biggr]dr
  \end{split}
\end{equation}
satisfies the identity
\begin{align*}
  (1-t)\partial_t E_{(q,p)}^{lm}(t)
  \le-\int_{S_t}\biggl[&(1+l(l+1))\left|\sqrt{1-t}f_{(q,p)}^{lm}\right|^2\\
    &+(1+(4(p-q)-3)t)\left|(1-t)\partial_tf_{(q,p)}^{lm}\right|^2\\
    &+\left|\sqrt{1-t}r \partial_r f_{(q,p)}^{lm}\right|^2 \biggr]dr\\
+\sqrt{1-t}\int_{S_t}\biggl[&
2(1-C_{p,q})\left(\sqrt{1-t}\bar f_{(q,p)}^{lm}\right)\left((1-t)\partial_tf_{(q,p)}^{lm}\right)\\
&+4(p-q-1)\left((1-t)\partial_t\bar f_{(q,p)}^{lm}\right)\left(\sqrt{1-t}\,r \partial_r f_{(q,p)}^{lm}\right)\\
&+\mathrm{c.\ c.}
\biggr]dr.
\end{align*}
The first integral (including the minus sign) can be bounded by $-E_{(q,p)}^{lm}(t)$ if $p-q\ge 1$ and by $E_{(q,p)}^{lm}(t)$ if $p-q=0$ (again the case $p-q<0$ is not of interest here). Using \emph{Young's inequality} of the form $a\overline b+\overline ab\le |a|^2+|b|^2$ for arbitrary complex numbers $a$ and $b$,
we can bound the second integral (including $\sqrt{1-t}$) on the right hand side by $C \sqrt{1-t}E_{(q,p)}^{lm}(t)$ for some unspecified constant $C>0$ (which in general depends on the choice of $p$ and $q$). Integrating the resulting inequality
\begin{equation*}
(1-t) \partial_t E_{(q,p)}^{lm}(t)\le -\kappa E_{(q,p)}^{lm}(t)+C \sqrt{1-t}E_{(q,p)}^{lm}(t),
\end{equation*}
with respect to time,
where $\kappa=1$ if $p-q\ge 1$ and $\kappa=-1$ if $p-q=0$,
we obtain the \emph{energy estimate}
\begin{equation}
    \label{eq:waveenergyest} 
    E_{(q,p)}^{lm}(t)\le C(1-t)^\kappa E_{(q,p)}^{lm}(0),
  \end{equation}
  for all $t\in [0,1)$ for a in general different constant $C>0$. Notice here that the definition of the energy in \Eqref{eq:waveenergydef} allows us to pull a common factor of $1-t$ out of the integral. Taking just the first term in the energy into account, the estimate \eqref{eq:waveenergyest} therefore suggests that $f_{(q,p)}^{lm}\sim O(1)$ at $t=1$ if $p-q\ge 1$ and $f_{(q,p)}^{lm}\sim O((1-t)^{-1})$ at $t=1$ if $p-q\ge 0$. The other two terms in the energy also give certain bounds for $f_{(q+1,p)}^{lm}$ and $f_{(q,p+1)}^{lm}$ which we ignore here.

We remark at this point that this energy estimate can actually be optimised slightly. This however requires a slightly more complicated expression for the energy. We do not discuss this here any further, but see \Sectionref{sec:nonlinearwaveenergy}.

As before for the spin-2 field we focus on Friedrich's strategy to split the field up into potentially singular contributions from the cylinder and fully regular contributions from the `bulk space-time'. To this end we pick arbitrary integers $\rho\ge 0$, $q_0\ge 0$ and $p_0\ge 1+q_0+\rho$ and use \Eqsref{eq:waveenergydef} and \eqref{eq:waveenergyest} so that 
\begin{align*}
  \sum_{p+q=p_0+q_0}^{p_0+q_0+\rho}\sum_{l=0}^\infty\sum_{m=-l}^l
\int_0^1\int_{S_t}(1+l(l+1))^{\rho+1}\left|f_{(q,p)}^{lm}\right|^2 dr dt\\
\le C \sum_{p+q=p_0+q_0}^{p_0+q_0+\rho}\sum_{l=0}^\infty\sum_{m=-l}^l (1+l(l+1))^{\rho} E_{(q,p)}^{lm}(0).
\end{align*}
where the right hand side (and therefore the left hand side) is guaranteed to be finite as a consequence of our fundamental smoothness assumptions. Since the left hand side bounds the square of the Sobolev norm of order $\rho$ of $\partial_t^{q_0}\partial_r^{p_0}f$, it follows
\[\partial_t^{q_0}\partial_r^{p_0}f\in H^\rho(D^+(r_*)\times\mathbb S^2),\]
and we conclude from the Sobolev embedding theorem that 
\[\partial_t^{q_0}\partial_r^{p_0}f\in C^{\rho-3}\left(\overline{D^+(r_*)\times\mathbb S^2}\right)\] 
 if $\rho\ge 3$. Again, what this means for us is that $\partial_t^{q_0}\partial_r^{p_0}f$ and all of its partial derivatives of order $\rho-3$ therefore extend continuously to the boundary of $D^+(r_*)\times\mathbb S^2$.
In particular, it follows that
\[\partial_r^{4}f\in C^{0}\left(\overline{D^+(r_*)\times\mathbb S^2}\right).\]
The identity 
\[f(t,r,x)=\sum_{p=0}^3\frac{1}{p!}\partial_{r}^pf(t,0,x)r^p+J^4[\partial_r^{4}f](t,r,x),\]
where $J$ is the map defined in \Eqref{eq:defJ}, therefore separates the field $f$ into a potentially singular contribution from the cylinder and a completely regular one from the bulk space-time,
as observed for the spin-2 field. More information about the potentially singular solutions of the intrinsic evolution equations on the cylinder $r=0$ are given in \Sectionref{sec:CWEMinkowski}.

As for the spin-2 system, let us remark briefly that the scalar wave equation has indeed a natural geometric energy generated by the standard scalar field energy-momentum tensor, see \cite{Szabados:2009ig}. We do not consider this energy here for similar reasons as for the spin-2 system.

\subsection{Some results for nonlinear wave equations}
\label{sec:nonlinearwaveenergy}

\renewcommand{\del}[1]{\partial_{#1}}

The only results regarding \emph{nonlinear} wave equations on cylinder-type space-times that we are currently aware of are from \cite{beyer2019a}. As discussed above, the main focus for nonlinear equations is the \emph{global-in-time existence problem} rather than sharp estimates for the asymptotic behaviour of the fields. The new small data global-in time existence result for Fuchsian-type PDEs  in \cite{oliynyk2016,beyer2019a} (with a further recent generalisation in \cite{fajman2020a}) is exploited in \cite{beyer2019a} to study a certain class of semi-linear wave equations on both the cylinder representations of Minkowski and the Schwarzschild space-time. In the interest of brevity we only give a very short summary of the results and compare them to those for the linear fields discussed above. Moreover, we simplify the presentation here by focussing on the spherically symmetric scalar case and on Minkowski space-time in the following. 

The class of semi-linear wave equations considered in \cite{beyer2019a} takes  the form
\begin{equation*}  
\bar g^{\alpha\beta}\bar\nabla_\alpha \bar\nabla_\beta \bar f = q(\bar f)\bar g^{\alpha\beta}\bar\nabla_\alpha \bar f \bar\nabla_\beta \bar f 
\end{equation*}
on the original (non-cylinder type) representation of Minkowski space-time
where $q \in C^\infty(\mathbb R^N)$. In the cylinder-representation of Minkowski space-time, this equation becomes
\begin{equation}
  \label{eq:nonlinwaveenergy}
  (1-t^2)\del{tt}f+2tr\del{tr}f-r^2\del{rr}f-2t\del{t} f-\Delta_{\mathbb S^2} f=g
\end{equation}
with
\begin{align*} 
g = -q\bigl(r(1-t)(1+t)f\bigr)\biggl(&2r(2t+(1-t)^2) f r\del{r}f
-r(1+t)^2 (1-t)^2(\del{t}f)^2\\
&+r(1+t)(1-t) r^2(\del{r}f)^2 + r(1+t)(1-t) f^2\\
&+2 r(1-3t-(1-t)^2)(1-t)(\del{t}fr\del{r}f-\del{t}f f)
\biggl).
\end{align*}
Because the nonlinear term $g$
satisfies the null condition of Klainerman\cite{klainerman1980}, global
existence results, under a small initial data condition, follow from the pioneering work in \cite{christodoulou1986}.
In this sense, the results here about \Eqref{eq:nonlinwaveenergy} are not new, but the method presented in \cite{beyer2019a} is. In particular the fact that the analysis has been carried out in Friedrich's cylinder representation gives a valuable new perspective. 

Defining $f_{(0,p)}=\partial_\rho^pf$ (time derivatives were not considered in \cite{beyer2019a}) for any integer $p\ge 0$, we choose 
\begin{align*}
    E_{(0,p)}(t)=(1-t)^{2\lambda-1}\int_{S_t}\Bigl[&(1+t)\left|(1-t)\partial_t f_{(0,p)}\right|^2
    +\left|\sqrt{1-t}\,r\partial_r f_{(0,p)}\right|^2 \\
  &+\left|f_{(0,p)}\right|^2\Bigr]dr
\end{align*}
as the \emph{energy},
  where $\lambda\in\mathbb R$ is so far an arbitrary constant. Given that \Eqref{eq:nonlinwaveenergy} is essentially the same kind of wave equation as \Eqref{eq:waveenergyswsh} (in the spherically symmetric case $l=0$), it is interesting to compare the energy here to the one in \Eqref{eq:waveenergydef} for $q=l=0$. Apart from the overall factor $(1-t)^{2\lambda-1}$, the two energies differ by a factor $1-t$ in the term $\left|f_{(0,p)}^{lm}\right|^2$. It turns out that this difference is not caused by the fact that the equation is nonlinear here and linear there. Both energies could be used in either case. In fact it turns out that the energy here would yield slightly better estimates also for the linear case. The reason why we have not used it in \Sectionref{sec:scalarwaveequationenergy} is that \Eqref{eq:waveenergydef} would become more complicated because the first term $(1+l(l+1))\left|\sqrt{1-t}f_{(q,p)}^{lm}\right|^2$ would need to be split up into two terms with different powers of $1-t$.

In any case, given this energy and the Fuchsian-type theorem of \cite{beyer2019a} (Theorem~3.8 there), it is possible to prove the following statement, skipping all of the technical details in our presentation here. Given \emph{sufficiently small} Cauchy data imposed at $t=0$ with sufficient regularity, the solution $f$ exists (within a fixed regularity class) on the whole time interval $[0,1)$. This therefore settles the global-in time existence problem. Moreover, provided the constant $\lambda$ in the definition of the energy is in $((2+\sqrt{2})/4,1]$, there exists a constant
$C>0$ such that
\[\sum_{p=0}^{p_0}E_p(t)\le C \sum_{p=0}^{p_0}E_p(0),\]
for some $p_0>0$. Using this energy estimate together with the definition of the energy above therefore yields that
\[\int_{S_t}\left|f_{(0,p)}\right|^2dr\le C (1-t)^{-\sqrt{2}/2},\]
for all $t\in[0,1)$ for some constant $C>0$. In comparison to the results for the linear case in \Sectionref{sec:scalarwaveequationenergy}, this estimate here is slightly better if $p=0$; our analysis in \Sectionref{sec:scalarwaveequationenergy} predicted $(1-t)^{-1}$ and this is purely a consequence of the slightly `better energy' used here. But the result here is worse than the one in \Sectionref{sec:scalarwaveequationenergy} whenever $p>0$. Open questions which are left for future research are therefore:
\begin{itemize}
\item Does differentiation with respect to $r$ make the solutions of \Eqref{eq:nonlinwaveenergy} more regular in the limit $t=1$ as it does for the linear wave equation as well as the spin-2 system? 
\item If yes, can we split the estimates as cleanly as in the previous linear cases into a `cylinder part' and a `bulk part'?
\item If yes, are potential singularities in the limit $t\rightarrow 1$ generated purely by the cylinder part, while the bulk is perfectly regular (as it is true for the linear cases before)? Or can the nonlinearities generate some additional singular bulk behaviour at $t=1$?
\end{itemize}
We shall not discuss the almost identical, but more technical, results (and open questions) regarding the same nonlinear wave equation on the cylinder-representation of the Schwarzschild space-time.


\renewcommand*{\d}{\mathrm{d}}
\renewcommand*{\del}{\partial}


\section{Numerical studies of fields near space-like and null infinity}
\label{sec:numer-stud-infinity}

\subsection{The Spin-2 equation on a Minkowski background\label{sec:spin2}}

Due to the previously discussed close relationship to the conformal field equations, the spin-2 system has been extensively numerically studied on various portions of Minkowski space \cite{Beyer:2012, Beyer:2014, Doulis:2013, Doulis:2017, PanossoMacedo:2018}. In the following, we review the most important results.

\subsubsection{First-order formulation near the cylinder}

The first investigation of the spin-2 equation, presented in \cite{Beyer:2012}, achieves the conformal compactification of Minkowski space and an appropriate representation of the cylinder with the following transformation due to Friedrich \cite{Friedrich:1998}. Starting from the physical metric $\tilde g$ in spherical coordinates $(\tilde t, \tilde r,\theta,\phi)$,
\begin{equation}\label{eq:MinkowskiMetric}
 \tilde g = \d\tilde t^{\,2}-\d\tilde r^2-\tilde r^2\d\omega^2,\quad \d\omega^2=(\d\theta^2+\sin^2\theta\,\d\phi^2),
\end{equation}
new coordinates $(t,r,\theta,\phi)$ in the region $\tilde r^2>\tilde t^2$ (the exterior of the null cone at the origin) are introduced according to
\begin{equation}
 r=\frac{\tilde r}{\tilde r^2-\tilde t^2},\quad
 t = \frac{\tilde t}{\tilde r\mu(r)}.
\end{equation}
Here, $\mu(r)$ is a smooth function with $\mu(0)=1$, which represents the remaining conformal gauge freedom. 
After rescaling the metric with the conformal factor
\begin{equation}
 \Theta = \frac{r}{\mu(r)}\left(1-t^2\mu^2(r)\right),
\end{equation}
one arrives at the conformal metric
\begin{equation}\label{eq:confmetric}
 g = \Theta^2\tilde g
   = \d t^2+2t\frac{\kappa'}{\kappa}\,\d t\,\d r-\frac{1-t^2\kappa'^2}{\kappa^2}\,\d r^2
     -\frac{1}{\mu^2}\,\d\omega^2,\quad
 \kappa(r):=r\mu(r),
\end{equation}
where $'$ denotes differentiation with respect to $r$.

\begin{figure}\centering
 \includegraphics[width=0.9\linewidth]{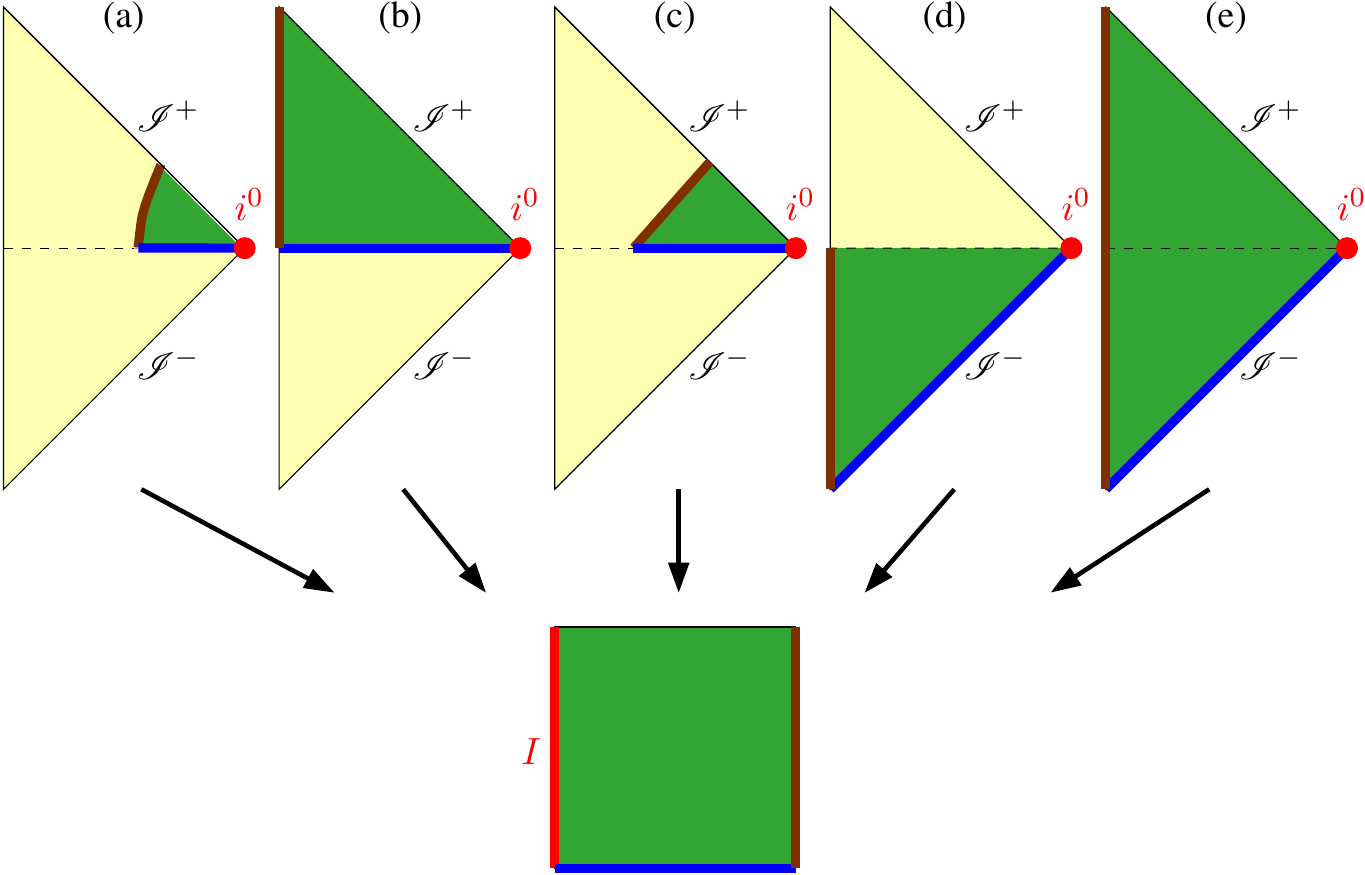}
 \caption{The spin-2 system (Sec.~\ref{sec:spin2}) and the conformally invariant wave equation (Sec.~\ref{sec:CWEMinkowski}) are studied on the subsets (a)-(e) of Minkowski space. In all cases, the relevant domain is mapped to a rectangular region. The cylinder $I$ is thereby blown up to the red boundary of the rectangle, whereas the `outer boundaries' of the numerical domains correspond to the brown lines. Initial data are prescribed at the blue surfaces.\label{fig:MinkowskiDomains}}
\end{figure}

In these coordinates, the cylinder $I$ is located at $r=0$, whereas the coordinate position of null infinity depends on the free function $\mu$. In the following, we consider the simplest (and, for numerical purposes, probably most suitable) choice  $\mu(r)\equiv 1$. In that case, $\scri^\pm$ are at $t=\pm 1$, respectively.\footnote{Note that, besides the choice $\mu=1$, also the case $\mu=1/(1+r)$ was studied in \cite{Beyer:2012}, for which $\scri^\pm$ are represented by straight lines with slopes $\pm1$. This turns out to have the disadvantage that the numerical investigations are restricted to a region that does not extend to $\scri^+$. Only the critical set $I^+$ is part of the numerical domain. Hence $\mu=1$ is the better choice if one is interested in the behaviour of the fields near  $\scri^+$.} 
For the numerical investigations, we focus on the domain $0\le t\le 1$, $0\le r\le 1$, which is illustrated in Fig.~\ref{fig:MinkowskiDomains}a.

As discussed in Sec.~\ref{sec:energyspin2}, in order to solve the spin-2 equation $\nabla^A{}_{A'} \phi_{ABCD} = 0$ on $g$, the five components $\phi_k$, $k=0,\dots, 4$, of the spin-2 zero-rest-mass field $\phi_{ABCD}$ are expanded into spin-weighted spherical harmonics $\Y{s}{lm}$ with $s=2-k$, $l\ge |s|$ and $|m| \le l$. This has the advantage that the equations for the modes $\phi_k^{lm}$ for different pairs $(l,m)$ decouple. From suitable linear combinations of the resulting equations, one obtains the five evolution equations (in the coordinates with $\mu\equiv 1$)
\begin{equation}\label{eq:spin2eqns}
  \begin{aligned}
(1+t) \del_t \phi^{lm}_0 - r\del_r \phi^{lm}_0 &= -2 \phi^{lm}_0   -\alpha_2 \phi^{lm}_1 ,\\
\del_t \phi^{lm}_1  &=  -  \phi^{lm}_1 +  \frac12 \alpha_2 \phi^{lm}_0 - \frac12 \alpha_0 \phi^{lm}_2,\\
\del_t \phi^{lm}_2  &=    \frac12 \alpha_0 \phi^{lm}_1 - \frac12 \alpha_0 \phi^{lm}_3,\\
\del_t \phi^{lm}_3  &=    \phi^{lm}_3 +  \frac12 \alpha_0 \phi^{lm}_2 - \frac12 \alpha_2 \phi^{lm}_4,\\
 (1-t) \del_t \phi^{lm}_4 + r\del_r \phi^{lm}_4 &=  2 \phi^{lm}_4 +  \alpha_2 \phi^{lm}_3,
  \end{aligned}
\end{equation}
and the three constraints
\begin{equation}\label{eq:Constraints}
  \begin{aligned}
  0 &= - 2 r \del_r \phi^{lm}_1 - 2 t \phi^{lm}_1 +  \alpha_0 (1-t) \phi^{lm}_2 + \alpha_2  (1+t) \phi^{lm}_0 ,\\
  0 &=-2 r\del_r \phi^{lm}_2   + \alpha_0 (1-t) \phi^{lm}_3 + \alpha_0 (1+t) \phi^{lm}_1,\\
   0 &=- 2 r \del_r \phi^{lm}_3 + 2 t \phi^{lm}_3 + \alpha_0 (1+t) \phi^{lm}_2 + \alpha_2    (1-t) \phi^{lm}_4,
  \end{aligned}
\end{equation}
where $\alpha_0=\sqrt{l(l+1)}$ and $\alpha_2=\sqrt{l(l+1)-2}$ (for $l\neq 0$) or $\alpha_2=0$ (for $l=0$). These are in fact (linear combinations of) the equations discussed in Sec.~\ref{sec:energyspin2}, cf.~\eqref{0evolswsh}-\eqref{3conswsh}.

The numerical problem then consists in finding suitable initial data at $t=0$ (the blue boundary in Fig.~\ref{fig:MinkowskiDomains}a) consistent with the constraints \eqref{eq:Constraints}, and evolving them in the region $0\le r\le 1$, $0\le t\le t_{\mathrm{max}}\le 1$ (the green region in Fig.~\ref{fig:MinkowskiDomains}a). Due to the property of the cylinder $I$ of being a totally characteristic surface, see Sec.~\ref{sec:sing-sets-behav}, no boundary conditions are needed there (i.e.\ at the red line in the figure). On the other hand, information \emph{can} travel through the boundary $r=1$ (the brown line), so that conditions are certainly required there for a well-posed mathematical problem. 

The first task of constructing initial data that satisfy the constraints can essentially be done by specifying two of the five functions at $t=0$ and using \eqref{eq:Constraints} to obtain the other three. As discussed in \cite{Beyer:2012}, a physically motivated procedure is to specify the `wave-like' degrees of freedom $\phi_0$ and $\phi_4$ and to solve the ODEs resulting from \eqref{eq:Constraints} at $t=0$ for $\phi_1$, $\phi_2$, $\phi_3$. Another, mathematically even more elegant way consists in specifying $\phi_2$ and $\phi_1-\phi_3$. Then all functions $\phi_k$ at $t=0$ can be obtained \emph{algebraically} from \eqref{eq:Constraints}, i.e.\ without solving ODEs. See \cite{Beyer:2012} for some further technical details.

The second task is to fix boundary conditions at the outer boundary $r=1$. It follows from the structure of the characteristics of \eqref{eq:spin2eqns} that only $\phi_0$ needs to be specified there, but none of the other functions. In a first numerical test in \cite{Beyer:2012}, which reproduces a known exact solution to the spin-2 system \cite{Doulis:2012}, the exact boundary values $\phi_0(t,1)$ were given as boundary condition. In other numerical examples, the condition $\phi_0(t,1)=0$ was chosen.

The evolution equations \eqref{eq:spin2eqns} are then numerically solved with the method of lines. For that purpose, the spatial derivatives are approximated with finite-differences based on particular summation by parts operators, and the resulting ODEs are solved with a standard fourth-order Runge-Kutta method. Thereby, an adaptive time-step is used, and the largest possible step is chosen that still satisfies the Courant-Friedrichs-Lewy (CFL) condition.
Finally, the boundary conditions are enforced with a penalty method, namely the simultaneous approximation term (SAT) method.

Note that the characteristic speed (corresponding to the $\phi_4$-equation) diverges as $t\to 1$, which forces the adaptive time step to become arbitrarily small near $\scri^+$. Consequently, the maximum time $t_\mathrm{max}$ needs to be strictly less than 1, and $\scri^+$ can only be approached but not reached exactly. In particular, a very large number of steps is required if $t_\mathrm{max}$ is close to 1.

The numerical experiments in \cite{Beyer:2012} show that the described method leads to reliable solutions to the spin-2 system. For example, for a certain choice of initial data and $400$ spatial gridpoints, the error (as measured with the $L^2$ norm) at $t=0.96$ in $\phi_0$  is in the order $10^{-4}$, whereas the error in $\phi_4$ (the quantity subject to the largest errors due to the diverging characteristic speeds) is around $10^{-2}$. Similarly, the constraints \eqref{eq:Constraints} are found to be initially satisfied very well, even though the errors increase near $\scri^+$.

Since the main motivation for these investigations is the desire to understand the behaviour of the fields near $I$ and $\scri^+$, it is somewhat unfortunate that the discussed method forces to choose $t_\mathrm{max}<1$. This issue was partially overcome in \cite{Beyer:2014} as follows. Initially, all fields were calculated up to the slice $t=t_\mathrm{max}$. Since the components $\phi_0$, $\phi_1$, $\phi_2$, $\phi_3$ all have `well-behaved' characteristics near $\scri^+$, it is possible to perform one Euler step to obtain values for these quantities at $t=1$. This does not work for $\phi_4$, though, due to the above-mentioned diverging characteristic speed. Instead, it can be used that the last evolution equation in \eqref{eq:spin2eqns} reduces to the intrinsic equation 
\begin{equation}\label{eq:phi4}
r\del_r \phi^{lm}_4 =  2 \phi^{lm}_4 +  \alpha_2 \phi^{lm}_3  
\end{equation}
on $\scri^+$. Regularity at $I^+$ ($t=1$, $r=0$) enforces the initial condition $\phi_4^{lm}(1,0)=-\frac{\alpha_2}{2}\phi_3^{lm}(1,0)$. Together with the information about $\phi_3$ on $\scri^+$, Eq.\ \eqref{eq:phi4} can be solved with this initial condition to finally also obtain the values of $\phi_4$ at $t=1$.

As discussed in \cite{Beyer:2014}, this does indeed allow one to produce quite accurate values of $\phi_0,\dots,\phi_3$ on $\scri^+$, provided one chooses sufficiently many spatial gridpoints. For example, the error in these quantities is in the order $10^{-7}$ for $3{,}200$ points. Unfortunately, the errors in $\phi_4$ are substantially larger: For the same example, they are around $0.1$. Hence this procedure can only provide rough approximations to the values on $\scri^+$.

\subsubsection{Second-order formulation near the cylinder}

The previous discussion of the spin-2 equation was based on a direct implementation of the first-order PDE system \eqref{eq:spin2eqns}. An interesting alternative approach is a reformulation as a system of wave equations. This second-order formulation, which was presented in \cite{Doulis:2013}, can be obtained by applying another spinor derivative $\nabla_{AA'}$ to the spin-2 equation $\nabla^{EA'}\phi_{EBCD}=0$. Using that the Weyl tensor of Minkowski space vanishes and that the chosen conformal metric has a vanishing scalar curvature, this leads to the wave equation
\begin{equation}\label{eq:waveEQ}
 \Box\phi_{ABCD}=0.
\end{equation}
The analysis in \cite{Doulis:2013} shows that the first- and second-order formulations are equivalent (have the same solutions), provided appropriate initial and boundary conditions for the wave equation are imposed. 

After a decomposition into spin-weighted spherical harmonics, the coordinate representation of \eqref{eq:waveEQ} reads (for $l\ge 2$)
\begin{equation}
 (1-t^2)\partial_{tt}\phi_k^{lm}-r^2\partial_{rr}\phi_k^{lm}+2tr\partial_{tr}\phi_k^{lm}
 +2(2-k-t)\partial_t\phi_k^{lm}+l(l+1)\phi_k^{lm}=0.
\end{equation}

This system can numerically be solved with the method of lines, similarly to the first-order problem. At the initial surface, function and time-derivative values need to be prescribed. After choosing the function values, the first-order system can be used to obtain the time-derivatives. At the outer boundary $r=1$, Robin conditions $\phi^{lm}_k(t,1)+\partial_r\phi^{lm}_k=g_k(t)$, $k=0,\dots 4$, are imposed. The source terms $g_k$ are read off from the constraints \eqref{eq:Constraints} (for $k=1,2,3$) and from the first-order evolution equations \eqref{eq:spin2eqns} (for $k=0, 4$).

The interesting result of numerical experiments is that the seemingly simple reformulation of the equation already reduces the error by a factor of about 4. Moreover, certain spurious high-frequency waves, which were observed in convergence plots for the first-order system, are not present in this approach. Hence the corresponding diagrams `look much cleaner'. These findings are in line with theoretical expectations according to which numerical approximations based on second-order equations generally have better properties than those for the corresponding first-order formulations \cite{Kreiss:2002a}.

\subsubsection{Global simulations}

A somewhat unsatisfactory feature of the above investigations of both the first- and second-order formulations of the spin-2 system is the `artificial' boundary at $r=1$, at which boundary conditions are required. While this leads to a mathematically perfectly valid problem, this boundary is `artificial' in the sense that there are no physical reasons why the functions should satisfy any particular known conditions there. Hence it is unclear whether one can obtain physically relevant solutions in this way. Instead, it is much more satisfactory to completely remove this boundary and solve the equations in the entire `upper half' of Minkowski space (the region shown in Fig.~\ref{fig:MinkowskiDomains}b). In that case, the brown boundary in the figure simply corresponds to the origin $\tilde r=0$, where no conditions apart from the usual regularity requirements are needed. This approach was presented in \cite{Doulis:2017}.

For that purpose, starting again from the Minkowski metric in spherical coordinates \eqref{eq:MinkowskiMetric}, one introduces null coordinates $u=\tilde t+\tilde r$, $v=\tilde t-\tilde r$, compactifies these by setting $p=\arctan u$, $q=\arctan v$, and then goes back to time-like and space-like coordinates $T=p+q$, $R=p-q$. Afterwards, the cylinder is `blown up' with the additional transformation $T=\kappa(r)F(t)$, $R=\pi r$, where the functions $\kappa$ and $F$ determine the shape and location both of the cylinder and of null infinity. 
Finally, after rescaling with the conformal factor $\Theta=\frac{2}{\kappa}\cos p\cos q$, the conformal metric becomes
\begin{equation}\label{eq:confmetric2}
 g=\Theta^2\tilde g
  =\dot F^2\d t^2 + \frac{2\kappa'F\dot F}{\kappa}\,\d t\,\d r
   -\frac{\pi^2-F^2\kappa'^2}{\kappa^2}\,\d r^2
   -\frac{\sin^2(\pi r)}{\kappa^2}(\d\theta^2+\sin^2\theta\,\d\phi^2),
\end{equation}
where $\dot{}$ and ${}'$ denote derivatives with respect to $t$ and $r$, respectively.
A possible choice of the free coordinate functions is $\kappa(r)=\cos(\pi r/2)$, $F(t)=\frac{1}{20}\ArcTanh(t)$, for which $I$ is located at $r=1$, while $\scri^\pm$ is (approximately, to machine accuracy) at $t=\pm1$.  

Since the previous numerical investigations highlight the advantage of second-order formulations, the considerations in \cite{Doulis:2017} are also based on a spin-2 wave equation rather than the first-order problem. For the present conformal metric, this equation reads
\begin{equation}
 \Box\Phi_{ABCD}+\frac{3\kappa^2}{4}\Phi_{ABCD}=0.
\end{equation}

Numerical solutions are then obtained with similar techniques as above, namely with the method of lines, based on suitable finite-difference approximations for the spatial derivatives, and time-integration with a fourth-order Runge-Kutta solver. The main difference to the previous numerical investigations is that certain regularity conditions need to be imposed at the origin $r=0$, in order to guarantee that formally singular terms in the equations (proportional to $1/r$ or $1/r^2$) are in fact regular. With these conditions, the limits of these terms can be worked out with L'H\^opital's rule. See \cite{Doulis:2017} for full technical details.

The numerical results show that, with an adaptive time-step, one can get very close to $\scri^+$. (A particular example with 600 spatial gridpoints managed to reach a final time $t_\mathrm{max}$ just $10^{-12}$ away from $t=1$.) The constraints were initially satisfied up to an error in the order $10^{-7}$, and, remarkably, the error did not grow substantially during the time-evolution. Hence quite accurate solutions can be obtained with this scheme. 

Note that this good accuracy is partially due to the choice of numerical examples that were studied in \cite{Doulis:2017}. We already discussed in Sec.~\ref{sec:sing-sets-behav} that certain logarithmic singularities can form on the cylinder as the critical set $I^+$ is approached, which are expected to `spread over' to $\scri^+$ and spoil the regularity there as well. This problem was avoided by choosing initial data at $t=0$ that rapidly go to zero as $I$ ($r=1$) is approached. Since these functions turn out to remain very close to zero near $I$ for $t>0$, no significant logarithmic singularities can form in these examples. On the other hand, for initial data that do not decay sufficiently quickly as $r\to1$, much less accurate results are expected with the discussed finite-difference methods. Therefore, we review an alternative approach in the next subsection.

\subsubsection{Spectral methods near the cylinder\label{sec:spectralspin2}}

The previous numerical investigations can successfully solve the spin-2 system and provide relatively accurate results. But it is not possible to reach $\scri^+$ exactly, and for optimal accuracy one needs to choose initial data for which logarithmic singularities are largely suppressed. However, since a detailed understanding of the behaviour of the solutions near $\scri^+$ and their regularity properties --- in particular in the presence of logarithmic singularities --- is one of the main motivations for carrying out such numerical simulations, it is desirable to consider methods that overcome these difficulties.

An important first step in this direction was made in \cite{PanossoMacedo:2018}, where a \emph{fully pseudospectral numerical method} was used, which is based on spectral expansions in terms of Chebyshev polynomials both in space \emph{and} time directions. This numerical scheme was first presented in \cite{Hennig:2009}, and some practical applications were studied in \cite{Ansorg:2011} and \cite{Hennig:2013}. Later the method was used to solve the conformally invariant wave equation on different background space-times \cite{Frauendiener:2014, Frauendiener:2017, Frauendiener:2018}, which we discuss in more detail in Secs.~\ref{sec:CWEMinkowski} and \ref{sec:CWESchwarzschild} below. But first we review the application to the spin-2 equation.

The starting point of the investigations in \cite{PanossoMacedo:2018} is the conformal metric \eqref{eq:confmetric} (again with the coordinate choice $\mu(r)\equiv 1$). However, instead of considering the green region in Fig.~\ref{fig:MinkowskiDomains}a, an additional coordinate transformation is used to map the region in Fig.~\ref{fig:MinkowskiDomains}c to a rectangle. The subtle but important difference is that the outer boundary (the brown line in the figure) now is a null geodesic. Hence no information can propagate through this boundary, and consequently no boundary conditions are required there. Alternatively to studying the entire upper half of Minkowski space (Fig.~\ref{fig:MinkowskiDomains}b), this is another elegant way to avoid artificial boundary conditions, which was already used in the above derivations of energy estimates in Sec.~\ref{sec:energyspin2}.

As before, the components of the spin-2 field are decomposed into spin-weighted spherical harmonics. Based on the expected behaviour of the solution near $I$ as derived with certain polylogarithmic expansions \cite{ValienteKroon2002}, the modes $\phi_k^{lm}$ are then further decomposed as (for $l\ge 2$)
\begin{equation}
 \phi_k^{lm}(t,r)=r^l [\alpha_k^{lm}(t)+r\beta_k^{lm}(t,r)],
\end{equation}
where the 1-dimensional $\alpha$-functions account for possible singularities, whereas the 2-dimensional $\beta$-functions are regular. According to the general theory, singular terms are expected to behave like $(1-t)^n\log(1-t)$ near $t=1$, where the exponent $n$ depends on $k$ and $l$. Since the evolution equations for $\alpha_k^{lm}$ and $\beta_k^{lm}$ decouple, thanks to the linearity of the spin-2 equations, one can separately study ODEs for the singular functions and PDEs for the regular functions.

The corresponding numerical investigations in \cite{PanossoMacedo:2018} show the well-known useful features of spectral methods. Not only can one obtain highly-accurate solutions (near machine accuracy), but the behaviour of the Chebyshev coefficients in the pseudospectral expansions directly reveals regularity properties of the solution. For analytic solutions, one obtains spectral convergence, where the Chebyshev coefficients decay exponentially. On the other hand, for solutions with $C^n$ regularity, the coefficients $c_j$ only fall off according to $c_j\sim j^{-(2n+3)}$. It turns out that the numerically observed decay rates for the coefficients of $\alpha_k^{lm}$ agree with the expected singular terms, whereas the Chebyshev coefficients for $\beta_k^{lm}$ are found to exhibit an exponential decay, which confirms the regularity of those functions.

A particularly important point in this approach is that $\scri^+$ is part of the computational domain. Hence, instead of only asymptotically approaching $\scri^+$ with a large number of time-steps, it can in fact directly be reached with a moderate number of gridpoints. This is related to the fact that effectively all gridpoints are coupled in the `highly implicit' fully pseudospectral scheme. Therefore, the physical domain of dependence is automatically contained in the numerical domain of dependence, and no restriction on the spacing via a CFL condition is required.

\subsection{The conformally invariant wave equation on a Minkowski background}
\label{sec:CWEMinkowski}

In the next two subsections, we consider another toy model for the conformal field equations, namely the conformally invariant wave equation on some background metric~$g$,
\begin{equation}\label{eq:CWE}
 g^{ab}\nabla_a\nabla_bf-\frac{R}{6}f=0,
\end{equation}
where $R$ is the Ricci scalar for $g$. This equation is conformally invariant in the sense that, if $f$ solves the equation for a metric $g$, then $\tilde f=\Theta^{-1}f$ solves it for the conformally rescaled metric $\tilde g=\Theta^2 g$.

While the \emph{scalar} equation \eqref{eq:CWE} is an even simpler model than the spin-2 system, it has the advantage that it can be studied for an arbitrary metric $g$. The spin-2 system, on the other hand, is restricted by an algebraic equation, the well-known Buchdahl condition. As a consequence, the spin-2 equation is inconsistent with space-times with nonvanishing Weyl tensor. Since there is no such limitation for the conformally invariant wave equation, this equation is particularly useful on curved backgrounds. Moreover, since \eqref{eq:CWE} has a similar structure of characteristics as the conformal field equations, a study of this equation should already provide interesting insights into properties and problems that can occur in the case of the full equations.

Before we look at a Schwarzschild background in the next subsection, we first summarise the results from the considerations for spherically symmetric solutions on a Minkowski space presented in \cite{Frauendiener:2014}. In that paper, the wave equation \eqref{eq:CWE} was solved with the fully-pseudospectral scheme described above. The numerical calculations were carried out on different portions of Minkowski space, namely the regions shown in Fig.~\ref{fig:MinkowskiDomains}b-e.

Firstly, a neighbourhood of the cylinder was considered as shown in Fig.~\ref{fig:MinkowskiDomains}c, exactly as in the above discussion of the spin-2 system with pseudospectral methods. This domain allows us to focus on a finite neighbourhood of the cylinder without the need of artificial boundary conditions at the outer boundary, which is an outgoing null geodesic. For the conformal metric \eqref{eq:confmetric}, the wave equation \eqref{eq:CWE} for spherically symmetric solutions takes the simple form
\begin{equation}
 (1-t^2)\del_{tt}f+2tr\del_{tr}f-r^2\del_{rr}f-2t\del_t f=0,
\end{equation}
which is \eqref{eq:waveenergy} without angular derivatives.
At the totally characteristic cylinder ($r=0$), this equation evidently becomes an intrinsic equation, which has the first integral $\partial_t f=d/(1-t^2)$ with some constant $d$. Hence initial data should be chosen subject to the regularity condition $\partial_tf=0$ at $t=0$, $r=0$, which guarantees regular solutions (with $d=0$). In the Schwarzschild case below we will see that an entire hierarchy of regularity conditions needs to be imposed for regularity of the solution at different orders. However, in the present Minkowski case, this single condition already guarantees regularity at all orders.

The numerical accuracy of the pseudospectral scheme is then tested by comparing the numerical results to several exact solutions. It is found that a moderate number of gridpoints (about 20-30 gridpoints both in space and time directions) is sufficient to produce highly-accurate numerical solutions (with absolute error around $10^{-12}$).

Secondly, the conformally invariant wave equation is globally solved in the region shown in Fig.~\ref{fig:MinkowskiDomains}b. This is done with the conformal metric \eqref{eq:confmetric2}, but with a different choice of the free functions than above. Here we choose $\kappa(r)=\cos r$, $F(r)=2t\kappa(r)$. A subsequent coordinate transformation then maps the upper half of Minkowski space to a unit square.

Furthermore, in addition to this global \emph{Cauchy} problem, two types of \emph{characteristic} initial value problems are investigated in \cite{Frauendiener:2014}. In both cases, initial data are given at $\scri^-$. These are then evolved either to the surface $\tilde t=0$ (Fig.~\ref{fig:MinkowskiDomains}d) or even up to $\scri^+$ (Fig.~\ref{fig:MinkowskiDomains}e), such that the entire Minkowski space is covered.

It turns out in numerical experiments that even these global considerations lead to highly-accurate numerical solutions. The only difference between simulations on the `bottom half' or the entire Minkowski space is that the numerical error is about 2-3 orders of magnitude smaller in the former compared to the latter case (in which the solution has to be computed on a much larger domain). However, in all examples the errors were below $10^{-10}$ even on the larger domain. 

Finally, we note a particularly interesting feature of the pseudospectral scheme. While in traditional time-marching schemes the error generally has an upward trend with increasing time, an analysis of the distribution of the errors over the numerical domain for the pseudospectral approach shows no such trend. Instead, the largest errors are observed near the symmetry axis. Apart from that, one finds errors of comparable sizes throughout the entire domain (cf.\ Fig.~6 in \cite{Frauendiener:2014}).

\subsection{The conformally invariant wave equation on a Schwarzschild background}
\label{sec:CWESchwarzschild}

After the successful investigations on a flat Minkowski space, it is an important question as to whether similar numerical methods can also handle a curved background. This was answered in the affirmative in \cite{Frauendiener:2017, Frauendiener:2018}, where it was demonstrated that the fully pseudospectral scheme does indeed produce highly-accurate numerical solutions on a Schwarzschild background as well. Moreover, similarly to the pseudospectral solution of the spin-2 system described in subsection \ref{sec:spectralspin2} above, the fall-off behaviour of the Chebyshev coefficients does then even provide extra information on the regularity properties of the solution.

\begin{figure}\centering
 \includegraphics[width=\linewidth]{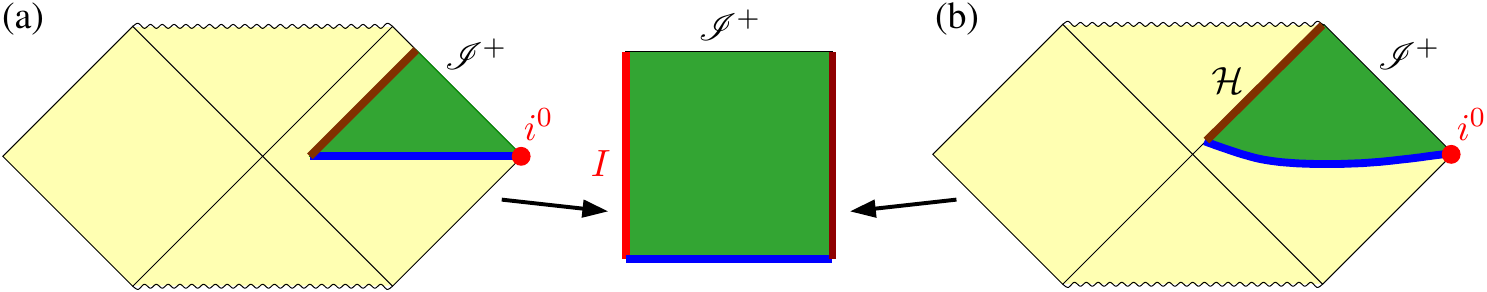}
 \caption{The conformally invariant wave equation on a Schwarzschild background is studied (a) in a neighbourhood of the cylinder $I$, and (b) in a domain that contains (a part of) the event horizon. Similarly to the considerations on a Minkowski background, the relevant domains are mapped to a rectangular region such that one side corresponds to $I$, initial conditions are given at the bottom, and $\scri^+$ is located at the top side.  \label{fig:SchwarzschildDomains}}
\end{figure}

The first numerical studies in \cite{Frauendiener:2017} consider {spherically-symmetric} solutions to the conformally invariant wave equation \eqref{eq:CWE}.
The underlying conformal compactification is obtained by starting from the Schwarzschild space-time in isotropic coordinates $(\tilde t,\tilde r,\theta,\phi)$ with the metric
\begin{equation}
 \tilde g=\left(\frac{1-\frac{m}{2\tilde r}}{1+\frac{m}{2\tilde r}}\right)^2
 \d {\tilde t}^{\,2}-\left(1+\frac{m}{2\tilde r}\right)^4[\d\tilde r^2+\tilde r^2(\d\theta^2+\sin^2\theta\,\d\phi^2)].
\end{equation}
Firstly, the transformation
\begin{equation}\label{eq:trans1}
 r=\frac{m}{2\tilde r},\quad t=\frac{2\tilde t}{m}.
\end{equation}
compactifies the radial coordinate (and introduces a dimensionless time coordinate). Secondly, the subsequent transformation
\begin{equation}\label{eq:trans2}
 t=\int_r^\rho\frac{\d s}{F(s)},\quad r=\rho(1-\tau),
  \quad\textrm{where}\quad
   F(s)=\frac{s^2(1-s)}{(1+s)^3},
\end{equation}
leads to coordinates $(\tau,\rho,\theta,\phi)$ in which space-like infinity is blown up. Finally, with the conformal factor
$\Theta=2r/[m(1+r)^2]$,
we obtain the conformal metric
\begin{equation}\label{eq:conmet}
 g=\Theta^2\tilde g=\frac{2}{\rho}A\,\d\rho\,\d\tau-\frac{1-\tau}{\rho^2}A[2-(1-\tau)A]\d\rho^2-\d\theta^2-\sin^2\theta\,\d\phi^2,
\end{equation}
where
\begin{equation}\label{eq:defA}
 A:=\frac{F(r)}{(1-\tau)^2F(\rho)}\equiv \frac{(1-r)(1+\rho)^3}{(1-\rho)(1+r)^3}.
\end{equation}
We choose the coordinate range $0\le\rho\le\rho_\mathrm{max}<1$ and $0\le\tau\le1$, which corresponds to the green region shown in Fig.~\ref{fig:SchwarzschildDomains}a. Note that the event horizon of the black hole is hidden behind a coordinate singularity at $\rho=1$. Hence, similarly to the Minkowski region considered in Fig.~\ref{fig:MinkowskiDomains}c, we focus a neighbourhood of the cylinder $I$ that contains a portion of $\scri^+$, and for which the outer boundary is a null geodesic at which no boundary conditions are required. In the present coordinates, the cylinder is located at $\rho=0$, and $\scri^+$ corresponds to $\tau=1$.

The conformally invariant wave equation then reads
\begin{equation}\label{eq:CWE1}
 (1-\tau)[2-(1-\tau)A]\,\del_{\tau\tau}f +2\rho\, \del_{\tau\rho}f
 -2\left[1-\frac{1-2r}{1-r^2}(1-\tau)A\right]\del_{\tau}f+\frac{4r A}{(1+r)^2}\,f=0.
\end{equation}
A systematic investigation of the analytical behaviour of the solution $f$ near the cylinder is carried out in \cite{Frauendiener:2017} by expanding the solution in the form $f(\tau,\rho)=f_0(\tau)+\rho f_1(\tau)+\rho^2 f_2(\tau)+...$ It turns out that, generally, there are singular terms at infinitely many orders $f_n$ at $\tau=1$. However, by appropriately restricting the initial data, one can eliminate arbitrarily many of the singular terms and achieve $C^k$ solutions for any $k$. The corresponding regularity conditions are algebraic restrictions on the initial data $f$ and $\del_\tau f$ and their $\rho$-derivatives at the `point' $I_0$ ($\rho=0$, $\tau=0$), where the initial surface $\tau=0$ intersects the cylinder.

The lowest-order singularity is present in $f_0$, but it disappears for data subject to $\del_\tau f=0$ at $I_0$. If this condition is imposed, then the next singular term appears in $f_2$ (while $f_1$ happens to be automatically regular), and can be eliminated by choosing data with $\del_{\tau\rho}f+\del_\rho f=0$ at $I_0$. Then, the next singularity in $f_3$ disappears if we choose data with $3\del_{\rho\rho}f+14\del_\rho f-36f=0$ at $I_0$. Similarly, arbitrarily many further singularities may be eliminated.

For numerical purposes, we need to impose at least the first regularity condition to remove the singularity in $f_0$, since otherwise the function values of $f$ themselves would diverge. The numerical experiments in \cite{Frauendiener:2017} show that, if we also satisfy one further regularity condition, then highly-accurate numerical solutions with absolute errors below $10^{-11}$ can be achieved. On the other hand, if only the first regularity condition is imposed, then the errors can grow up to about $10^{-6}$. While this is still perfectly fine in comparison to finite-difference schemes, it is not quite the accuracy that one usually achieves with (pseudo)spectral methods. However, we will see  below that the high accuracy can be restored with the help of an additional coordinate transformation.

Finally, it is observed that the error decays algebraically rather than exponentially, as expected, corresponding to the finite $C^k$ regularity of the solutions.

While these considerations are based on the computational domain shown in Fig.~\ref{fig:SchwarzschildDomains}a, which does not contain the event horizon, it is possible to modify the conformal compactification and consider the region in Fig.~\ref{fig:SchwarzschildDomains}b, which does extend to the horizon. In that case, however, one needs to restrict the numerical calculations to a time interval $0\le\tau\le\tau_{\mathrm{max}}<1$ bounded away from $\scri^+$, as otherwise future time-like infinity $i^+$ would be part of the computational domain. Since this point is singular in the Schwarzschild space-time, the solutions to the conformally invariant wave equation are generally singular there as well. On that restricted time interval, the numerical calculations can be carried out and lead to highly-accurate solutions. One even observes spectral convergence, since the logarithmic singularities are then outside the numerical domain. See \cite{Frauendiener:2017} for more details.

So far we have restricted ourselves to spherically-symmetric solutions, for which~$f$ depends on $\tau$ and $\rho$ only, but it is straightforward to extend the investigations to nonspherical waves \cite{Frauendiener:2018}. By expanding $f$ into spherical harmonics, one obtains equations for the different modes. It turns out that each mode $f_{lm}$ almost satisfies \eqref{eq:CWE1}, we just need to include the additional term $Al(l+1)f_{lm}$ on the left-hand side. 

Similarly to the spherically-symmetric situation above (corresponding to $l=0$), the behaviour near the cylinder can also be studied in the general case. For each $l$, one obtains a hierarchy of regularity conditions that enforce increasing degrees of regularity of the corresponding mode. 

The pseudospectral scheme works for the nonspherical waves too. Again it is found that sufficiently regular solutions (corresponding to initial data satisfying the first few regularity conditions) can be calculated with very high accuracy, while the errors are larger for less regular solutions. However, it is also demonstrated in \cite{Frauendiener:2018} that one can easily improve the regularity in these cases. By introducing a new time coordinate $t$ via $\tau=1-\ee^{-\frac{wt}{1-t}}$ and an appropriate choice of the parameter $w$ (usually values around $w=3$ turn out to work well), steep gradients of the solution near $\tau=1$ can be reduced. Effectively, this transformation converts $C^k$ functions of $\tau$ into $C^\infty$ functions of $t$. This achieves that the numerical error is reduced to values below $10^{-11}$ even for solutions with singularities at low orders.

Probably the most important result in \cite{Frauendiener:2018} concerns the regularity of solutions near $\scri^+$. The analytical investigation of the intrinsic equations at the cylinder ($\rho=0$) show the presence of logarithmic singularities in $f$ and its $\rho$-derivatives at the cylinder as $\tau\to1$, but this does not reveal how the solutions behave if $\tau=1$ is approached along some other line $\rho=\mathrm{constant}\neq 0$ away from the cylinder. One would expect that singularities at the critical set $I^+$ spread to $\scri^+$ and that similar logarithmic terms should generally be present near $\tau=1$. Numerically, this is tested by observing the fall-off behaviour of the Chebyshev coefficients, and it is found that the solutions do show the expected $C^k$ behaviour at $\scri^+$. This provides strong numerical evidence that singularities indeed spread from the cylinder via the critical set to $\scri^+$. While this was specifically observed for the relatively simple conformally invariant wave equation, it is expected that solutions to the full Einstein equations generically share the property of a limited regularity and appearance of logarithmic singularities both at the cylinder and at $\scri^+$.

\subsection{Future studies}

The encouraging results in the successful investigations of the spin-2 system and the conformally invariant wave equation on Minkowski or Schwarzschild backgrounds show that the behaviour of fields near space-like and null infinity can be simulated with very good numerical accuracy. This should motivate various future investigations, for example computations on less symmetric background space-times and studies of model equations that go beyond linear approximations. The ultimate goal, of course, would be simulations based on the full conformal field equations.

A starting point could be the conformally invariant wave equation on a Kerr background. We conclude this section with a short outlook on corresponding work in progress and show how a suitable conformal compactification of the Kerr space-time can be obtained \cite{HennigPanossoMacedo}. Appropriate coordinates should (i) cover a part of space-like and future null infinity, (ii) `blow up' space-like infinity to the cylinder $I$, (iii) have the property that an outer boundary at constant radial coordinate is a null surface through which no information can travel, in order to avoid artificial boundary conditions. This can be done as follows \cite{HennigPanossoMacedo}.

Starting from the Kerr metric in Boyer Lindquist coordinates $(\tilde r,\theta,\varphi,\tilde t)$,
\begin{equation}
 \tilde g = \tilde\Sigma\left(\frac{\d\tilde r^2}{\tilde\Delta}+\d\theta^2\right)
             +(\tilde r^2+a^2)\sin^2\theta\,\d\varphi^2
             -\d\tilde t^2+\frac{2m\tilde r}{\tilde\Sigma}(a\sin^2\theta\,\d\varphi-\d\tilde t)^2
\end{equation}
with $\tilde\Sigma=\tilde r^2+a^2\cos^2\theta$ and $\tilde\Delta=\tilde r^2-2m\tilde r+a^2=:(\tilde r-\tilde r_+)(\tilde r-\tilde r_-)$, we first perform a transformation $\varphi\mapsto\phi$,
\begin{equation}\label{eq:phitrans}
 \d\varphi=\d\phi+\frac{a}{\tilde\Delta}\,\d\tilde r.
\end{equation}
Then we compactify $\tilde r$ and define dimensionless coordinates $r$ and $t$,
\begin{equation}
 \tilde r = \frac{\tilde r_+}{r},\quad \tilde t=\tilde r_+ t.
\end{equation}
Next it is useful to introduce $\kappa:=a/\tilde r_+$ and $\tilde r_+$ as the fundamental parameters. Finally, we compactify the time coordinate and blow up $i^0$ in such a way that the new coordinate $\rho$ is a null coordinate. Based on the study of a family of null geodesics, this turns out to be possible with the transformation
\begin{equation}
 r=\rho(1-\tau),\quad
 t=\int_r^\rho \frac{\d s}{F(s)},
\end{equation}
where 
$F(r):=r^2\Delta/(1+\kappa^2 r^2)$ and
$\Delta:=\frac{r^2}{\tilde r^2_+}\tilde \Delta\equiv(1-r)(1-\kappa^2 r)$.
If we choose the conformal factor $\Theta=r/\tilde r_+$ and define
$\Sigma:=\frac{r^2}{\tilde r^2_+}\Sigma\equiv1+\kappa^2 r^2\cos^2\theta$, then the conformal metric $g=\Theta^2\tilde g$ has the following components in the coordinates $(\rho,\theta,\phi,\tau)$,
\begin{equation*}
 \newcommand{\squeeze}{}
 \left(\begin{array}{cccc}
    \frac{1}{F(\rho)}\left(\frac{(1+\kappa^2)r-\Sigma}{\Sigma F(\rho)}r^2+2(1-\tau)\right)
    & \squeeze 0
    & \squeeze -(1-\tau+\frac{(1+\kappa^2)r^3}{\Sigma F(\rho)})\kappa\sin^2\theta
    & \squeeze -\frac{\rho}{F(\rho)}\\
    0 & \squeeze \Sigma & \squeeze 0 & \squeeze 0\\
    -(1-\tau+\frac{(1+\kappa^2)r^3}{\Sigma F(\rho)})\kappa\sin^2\theta
    & \squeeze 0
    & \squeeze \left(1+\kappa^2 r^2+\frac{(1+\kappa^2)r^3 \kappa^2\sin^2\theta}{\Sigma}\right)\!\sin^2\theta
    & \squeeze \kappa\rho\sin^2\theta\\
    -\frac{\rho}{F(\rho)}
    & \squeeze 0
    & \squeeze \kappa\rho\sin^2\theta
    & \squeeze 0
   \end{array}\right)\!.
\end{equation*}
In these coordinates, curves of the form $\rho=\mathrm{constant}$, $\theta=\mathrm{constant}$, $\phi=\mathrm{constant}$, $t=\lambda$ are null geodesics of $g$. A crucial point in this compactification was the initial transformation of the angle $\varphi$, which effectively achieves that the above function $F(r)$ is independent of $\theta$.

The next step will be the numerical solution of the conformally invariant wave equation on this background. Particularly interesting is the question as to what extend the rotating Kerr background changes the picture from the Schwarzschild analysis. An important technical difference is the fact that the angular dependence cannot be separated in the Kerr case. Consequently, we need to solve an at least $2+1$ dimensional problem (if we restrict ourselves to axisymmetric waves). Note that the fully pseudospectral scheme can handle this type of problem as well \cite{PanossoMacedo:2014}. Corresponding numerical investigations will be presented elsewhere \cite{HennigPanossoMacedo}.


\section*{Acknowledgments}
This research was funded in part by the Marsden Fund of the Royal Society of New Zealand under grant numbers UOO0922 and UOO1924 and by a grant from the Division of Sciences of the University of Otago.


\def\BIBand{and}

\end{document}